\documentclass[twoside,twocolumn]{article}

\usepackage{amssymb,graphicx,url,amsmath,subfigure
}

\usepackage{verbatim}
\usepackage{times}
\usepackage{tikz}
\usepackage{soul}
\usepackage{color}
\usepackage{xcolor}
\usepackage{algorithm}
\usepackage{algorithmic}
\usepackage{enumerate}
\usepackage{comment}

\usepackage{etoolbox}
\makeatletter
\patchcmd{\Ginclude@eps}{"#1"}{#1}{}{}

\usepackage{amsthm}

\usetikzlibrary{shapes.geometric,positioning}

\newcommand{\cA}{{\cal A}}

\newcommand{\R}{{\mathbb R}}

\newcommand{\cX}{{\cal X}}
\newcommand{\cY}{{\cal Y}}

\newcommand{\bX}{{\bf X}}

\newcommand{\bK}{{\bf K}}

\newcommand{\bk}{{\bf k}}

\newcommand{\bN}{{\bf N}}

\newtheorem{Theorem}{Theorem}

\newtheorem{Definition}{Definition}
\newtheorem{Example}{Example}

\newtheorem{Question}{Question}

\newtheorem{Proposition}{Proposition}

\newtheorem{Idea}{Idea}

\newenvironment{Proof}{Proof:}{$\Box$}

\newcommand\independent{\protect\mathpalette{\protect\independenT}{\perp}}
\def\independenT#1#2{\mathrel{\rlap{$#1#2$}\mkern2mu{#1#2}}}

\definecolor{MyDarkGreen}{rgb}{0.17,0.46,0.25} 
\definecolor{MyDarkRed}{rgb}{0.88,0.22,0.21} 
\definecolor{MyDarkBlue}{rgb}{0.11,0.11,0.70}

\definecolor{lightgray}{gray}{0.85}
\sethlcolor{Dandelion}

\usetikzlibrary{arrows,shapes,plotmarks,positioning}
\usetikzlibrary{decorations.markings}

\tikzset{>=stealth'} 
\tikzstyle{graphnode} = 
 [circle,draw=black,minimum size=22pt,text centered,text
 width=22pt,inner sep=0pt] 
\tikzstyle{var} =[graphnode,fill=white]
\tikzstyle{vardashed} =[graphnode,draw=gray,fill=white]
\tikzstyle{obs} =[graphnode,fill=black,text=white]
\tikzstyle{obsgrey} =[graphnode,draw=white,fill=lightgray,text=black]
\tikzstyle{par} =[graphnode,draw=white,fill=red,text=black] 
 \tikzstyle{crucial} =[graphnode,draw=white,fill=yellow,text=black] 
\tikzstyle{fac} =[rectangle,draw=black,fill=black!25,minimum size=5pt]
\tikzstyle{facprior} =[rectangle,draw=black,fill=black,text=white,minimum size=5pt]
\tikzstyle{edge} =[draw=white,double=black,very thick,-]
\tikzstyle{blueedge} =[draw=white,double=blue,very thick,-]
\tikzstyle{rededge} =[draw=white,double=red,very thick,-]
\tikzstyle{prior} =[rectangle, draw=black, fill=black, minimum size=
5pt, inner sep=0pt]
\tikzstyle{dirprior} = [circle, draw=black, fill=black, minimum
size=5pt, inner sep=0pt]

\tikzstyle{dot_node}=[draw=black,fill=black,shape=circle]

\date{15 November 2022}

\begin{document}
\frenchspacing

\title{{\bf Phenomenological Causality}}

\author{Dominik Janzing$^1$  and Sergio Hernan Garrido Mejia$^{1,2}$\\
{\small 1) Amazon Research T\"ubingen, Germany, 2) Max Planck Institute for Intelligent Systems, Tübingen, Germany}\\
{\small janzind@amazon.com, shgm@tuebingen.mpg.de }}

\maketitle

\begin{abstract}
Discussions on causal relations in real life often consider variables for which
the {\it definition} of causality is unclear since the notion of {\it interventions} on the respective variables is obscure. 
Asking `what qualifies an action for being an intervention on the variable $X$' 
raises the question whether the action impacted all other variables only {\it through $X$} or {\it directly}, which implicitly refers to a causal model.   

To avoid this known circularity, we instead suggest a notion of `phenomenological causality' whose basic concept is a set of {\it elementary actions}. Then the causal structure is defined such that elementary actions change only the causal mechanism   at {\it one} node (e.g. one of the causal conditionals in the Markov factorization). 
This way, the Principle of Independent Mechanisms becomes the defining property of causal structure in domains where 
causality is a more abstract phenomenon rather than being an objective fact relying on hard-wired causal links between tangible objects.   
We describe this phenomenological approach to causality for toy and hypothetical real-world examples and argue that
it is consistent with the causal Markov condition when the system under consideration interacts with other variables that control the elementary actions. 

\end{abstract}

\section{Introduction} 

While machine learning (ML)  plays an increasing role in the technological development of our society (for instance,  
by providing forecasting systems, search engines, recommendation systems, and logistic planning tools) the ability of ML systems to learn {\it causal}  structures as opposed to mere statistical associations 
is still at an early stage\footnote{Cf. for instance, \cite{Pearl2018}, page 30: ``Some readers may be surprised that I placed present-day learning machines squarely on rung one of the Ladder of Causation...''}. A particularly hard task in causal machine learning is so-called {\it causal discovery}, the task of learning the causal graph (including causal directions) from passive observations
\cite{Spirtes1993,Pearl:00,causality_book}. Given how many problems modern deep learning (DL) has solved via providing computers with massive data \cite{d2l}, one may wonder whether appropriate DL architectures could {\it learn how to learn} causal structure from data 
after feeding them with an abundance 
of datasets with known ground truth. This approach, however, fails alone due to the scarcity of such datasets.\footnote{Even benchmarking the elementary problem of cause-effect inference from bivariate data is often done via the Tübingen dataset \cite{cepairs} containing currently 106 pairs only \cite{Guyon2019}. A lot odf studies are therefore heavily based on simulated data \cite{Ke2020}.} 
This, in turn, raises the question: {\it why is there so little benchmarking data with commonly agreed causal structure?}
The common answer is that in many real world systems interventions are impossible, costly, or unethical.
(e.g. moving the moon to show that its position causes the solar eclipse is costly at least). 
Although this is a valid explanation, it blurs the question whether the required interventions are {\it well-defined} in the first place.
While {\it defining} interventions on the moon seems unproblematic, it will be significantly harder to agree on a definition for interventions on 
the Gross National Product (GNP)', for instance, as a basis for discussing the impact of GNP on employment. --  
Which of all hypothetical political instruments (whether feasible or not) influencing GNP should be considered {\it interventions on GNP}? 
Before going into this discussion, we first recall the description of interventions in the framework of graphical models. 

   
\paragraph{Notation and terminology}   
Random variables will be denoted by capital letters like $X,Y$ and their values by lower case letters $x,y$.   
Further, calligraphy letters like $\cX,\cY$ will denote the range of variables $X,Y$.   
Causal Bayesian Networks (CBNs) or 
Functional Causal Models (FCMs) \cite{Spirtes1993,Pearl:00} will be our key framework for further discussions. 
Both concepts describe causal relations between
random variables $X_1,\dots,X_n$ via directed acyclic graphs (DAGs). According to the Causal Markov Condition, the respective causal DAG $G$ is compatible with any joint density that factorizes\footnote{Here we have implicitly assumed that the joint distribution has a density with respect to a product measure \cite{Lauritzen}.} according to
\begin{equation}\label{eq:fac}
p(x_1,\dots,x_n) = \prod_{j=1}^n p(x_j|pa_j), 
\end{equation} 
where $pa_j$ denotes the values of the parents $PA_j$ of $X_j$ in $G$. Then, the CBN is given by a DAG $G$ together with a compatible joint distribution. 

FCMs, in contrast, provide a deterministic model of the joint distribution, in which every node is a function of its parents 
and an unobserved noise variable $N_j$:
\begin{equation}\label{eq:fcm} 
X_j = f_j(PA_j,N_j),
\end{equation}
where all $N_1,\dots,N_n$ are statistically independent. Both frameworks, CBNs and FCMs
admit the derivation of interventional probabilities, e.g., the change of the joint distribution after setting variables to fixed values. While CBNs only provide statements on how probabilities change by an intervention, FCMs, in addition, tell us how the intervention affected each individual statistical unit and counterfactual causal statements  \cite{Pearl:00}. 

\paragraph{Interventions}
To briefly review different types of interventions, note that
the point intervention $do(X_j=x_j)$  \cite{Pearl:00}, also called `hard intervention', adjusts the variable to  $x_j$, while generalized (also called `soft') interventions on $X_j$ replace 
$p(x_j|pa_j)$ with a different conditional $\tilde{p}(x_j|pa_j)$ or the FCM \eqref{eq:fcm} with a modification $X_j = \tilde{f}_j(PA_j,\tilde{N}_j)$, see
 \cite{causality_book}, page 89, and references therein. Structure-preserving interventions \cite{CIC2020} preserve all the dependences on the parents 
 by either operating on the noise $N_j$ or adjusting $X_j$ to the parent-dependent value $f_j(pa_j,N'_j)$ where $N_j'$ is an independent copy of $N_j$ that is generated by the experimenter. The simple observation that any of these interventions on $X_j$ affect only $X_j$ and a subset of its descendants,
defines a consistency condition between a hypothetical $G$ and the hypothesis that an action is an intervention on $X_j$.

While the above framework has provided a powerful language for a wide range of causal problems, 
it implicitly requires the following two questions to be clarified:
\begin{Question}[coordinatization]\label{q:vardef} 
How do we define variables $X_1,\dots,X_n$ for a  given system, that are not only meaningful in their own right but also 
allow for well-defined causal relations between them?
\end{Question}
The second question reads: 
\begin{Question}[defining interventions] \label{q:intervention} 
Let $A$ be an intervention on a system $S$ whose state is described
 by the variables $\{X_1,\dots,X_n\}$ (`coordinates').
What qualifies $A$ to be an intervention on variable $X_j$ only?
\end{Question} 
Note that the word `only' in Question \ref{q:intervention} is meant in the sense that the action intervenes on none of the other variables under consideration 
{\it directly}, it only affects them in their role of descendants of $X_j$, if they are. 

Question 1 captures what is often called `causal representation learning'  \cite{Chalupka2015,Schoelkopf2021}, while this paper will 
mainly focus on Question 2 only although we believe that future research may not consider them as separate questions for several reasons.
First, because some definitions of variables may seem more meaningful than others because they admit more natural definitions of interventions. 
Second, because different parameterizations of the space results in different {\it consistency conditions} between variables which may or may not be
interpreted as causal interactions between them.  

Following e.g. \cite{Woodward2003} page 98, 
one may define interventions on $X_j$ as actions that affect only $X_j$
and its descendants, but this way one runs into the circularity of referring to the causal structure for defining interventions although one would like to define {\it causality via interventions}, see e.g. \cite{Baumgartner2009} for a discussion.       
The idea of this paper is to define causal directions between a set of variables by first defining a set of {\it elementary transformations}
acting on the system, which are later thought of being interventions on one of the variables only. While they can be concatenated to more complex transformations, defining which transformations are {\it elementary}, defines the causal direction. 
While the notion of interventions comes as a {\it primary} concept in the graphical model based framework of causal inference this paper will tentatively describe a notion of intervention as a {\it secondary} concept derived from an implicit or explicit notion of
{\it complexity of actions}.


\paragraph{Structure of the paper} 
Section \ref{sec:hardware} argues that there are domains (mostly technical devices) in which an action can be identified as an intervention on a certain variable via 
analyzing the `hardware' of the respective system.
In contrast, Section \ref{sec:ill} describes a few scenarios with ill-defined causal relations to highlight the limitations of the `hardware analysis' approach, without claiming that our proposal will offer a solution to all of them.   
Section \ref{sec:im} argues that the idea that some operations are more elementary than others is already necessary to make sense
of one version to read the Principle of Independent Mechanisms.
Based on this insight, Section \ref{sec:element}, which is the main part of the paper, describes how to define causal directions via declaring a set of transformations as elementary
and illustrates this idea for examples of `phenomenological' causality where the causal directions are debatable and paradoxical. 
Section \ref{sec:coupling}  shows that phenomenological causality appears more natural when the system under consideration is not considered in isolation,
but in the context of further variables in the world. Then, the joint system can be described by a DAG that is consistent with phenomenological causality
(Subsection \ref{subsec:MarkovExt}). Further, the notion of `elementary' also satisfies a certain consistency condition with respect to such an extension (Subsection \ref{subsec:boundary}).

\section{Defining interventions via `hardware analysis‘ \label{sec:hardware}}  
\begin{figure} 
\includegraphics[width=0.5\textwidth]{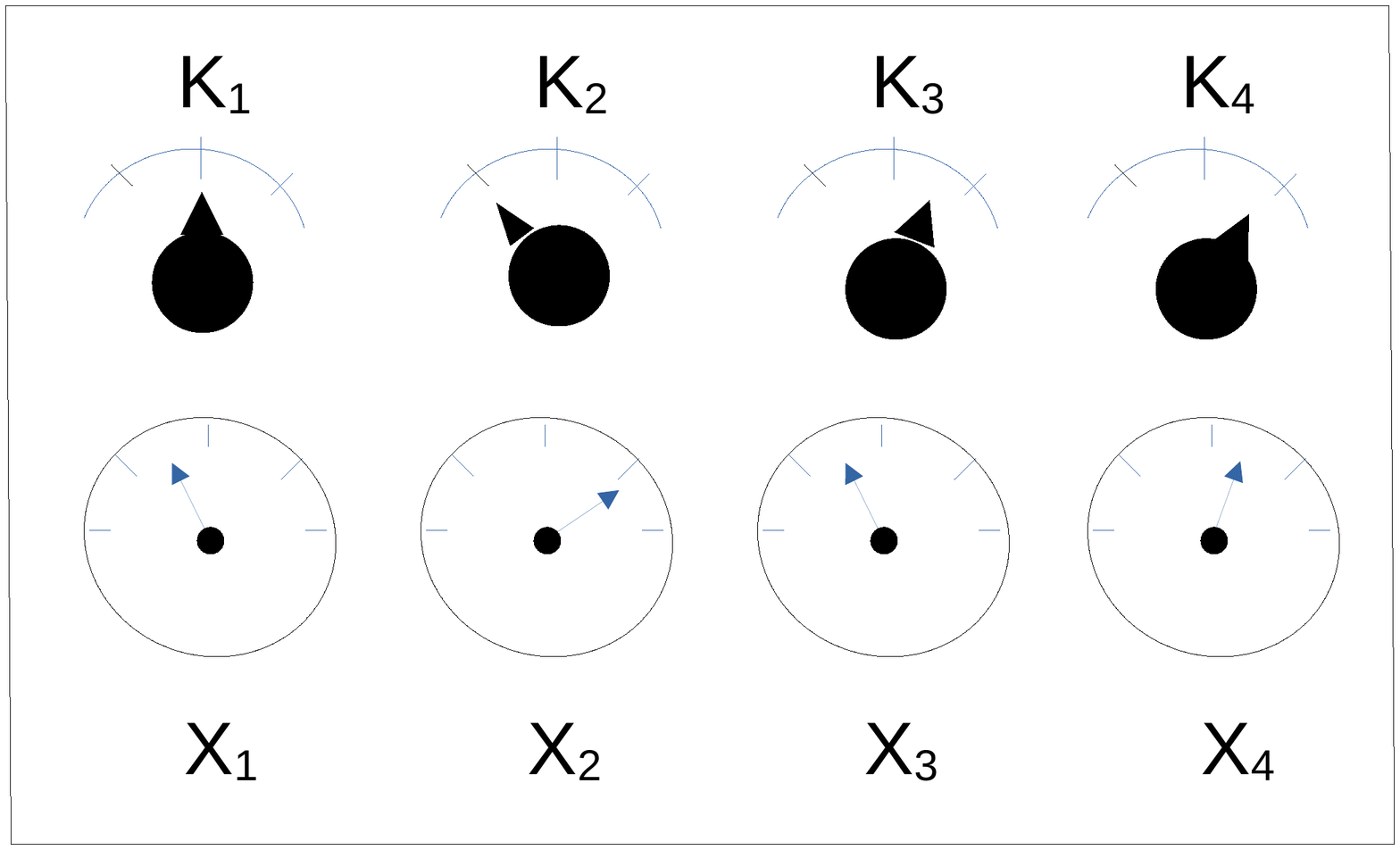}
\caption{\label{fig:apparatus} Apparatus whose front side contains $n$ measurement devices and $n$ knobs. The measuring devices measure 
unknown quantities $X_1,\dots,X_n$. How do we `find out' (or `define'?) whether knob $j$ intervenes on $X_j$?} 
\end{figure}

We want to motivate Question 1 and Question 2 from the previous section by two thoughts experiments, starting with Question \ref{q:intervention}. 
Consider the apparatus shown in figure \ref{fig:apparatus}.
 It shows $n$ measuring devices that display real numbers 
$X_1,X_2,\dots,X_n$. Further, it contains  $n$ knobs, whose positions are denoted by $K_1,\dots,K_n$. Assume we know 
that the box contains an electrical device and $X_j$ are $n$ different voltages whose mutual influence is described by some unknown DAG.
In the best case, our knowledge of how the device is internally wired tells us that turning knob $j$ amounts to intervening on $X_j$. 
In the worst case (the black box scenario) our judgement of whether $K_j$ intervenes on $X_j$ is only based on observing the impact on all $X_i$, where we run in the above mentioned circularity of checking whether $K_j$ only affects $X_j$ and a subset of its descendants. 
However, depending on `hardware analysis' for defining interventions in a non-circular way is worrisome. 
First, the power of the causal inference framework relies on the fact that it describes causal relations on a more abstract level without referring to the underlying `hardware'. Second, it is questionable why analyzing the causal relation between the action at hand and a variable $X_j$ should be easier
than analyzing the causal relations between different $X_i$ (e.g. following wires between the knobs and the voltages $X_j$ as well as
the wires between different $X_i$ both requires opening the box). After all, both causal questions refer to the same domain. 
A large number of relevant causal relations refer to domains with an inherent fuzziness, e.g., macro economic questions, and it is likely that the same 
fuzziness applies to the causal relation between an action and the variables it is supposed to intervene or not to intervene on. 
Variables for technical devices like {\it voltage at a specific component} refer to measurements that are local in space-time and thus require propagating signals 
to interact, which admits interpreting edges of a causal DAG as those signals. 
Coarse-grained variables like GNP are highly non-local, which renders causal edges an abstract concept. 

\begin{figure}[h]  
\includegraphics[width=0.5\textwidth]{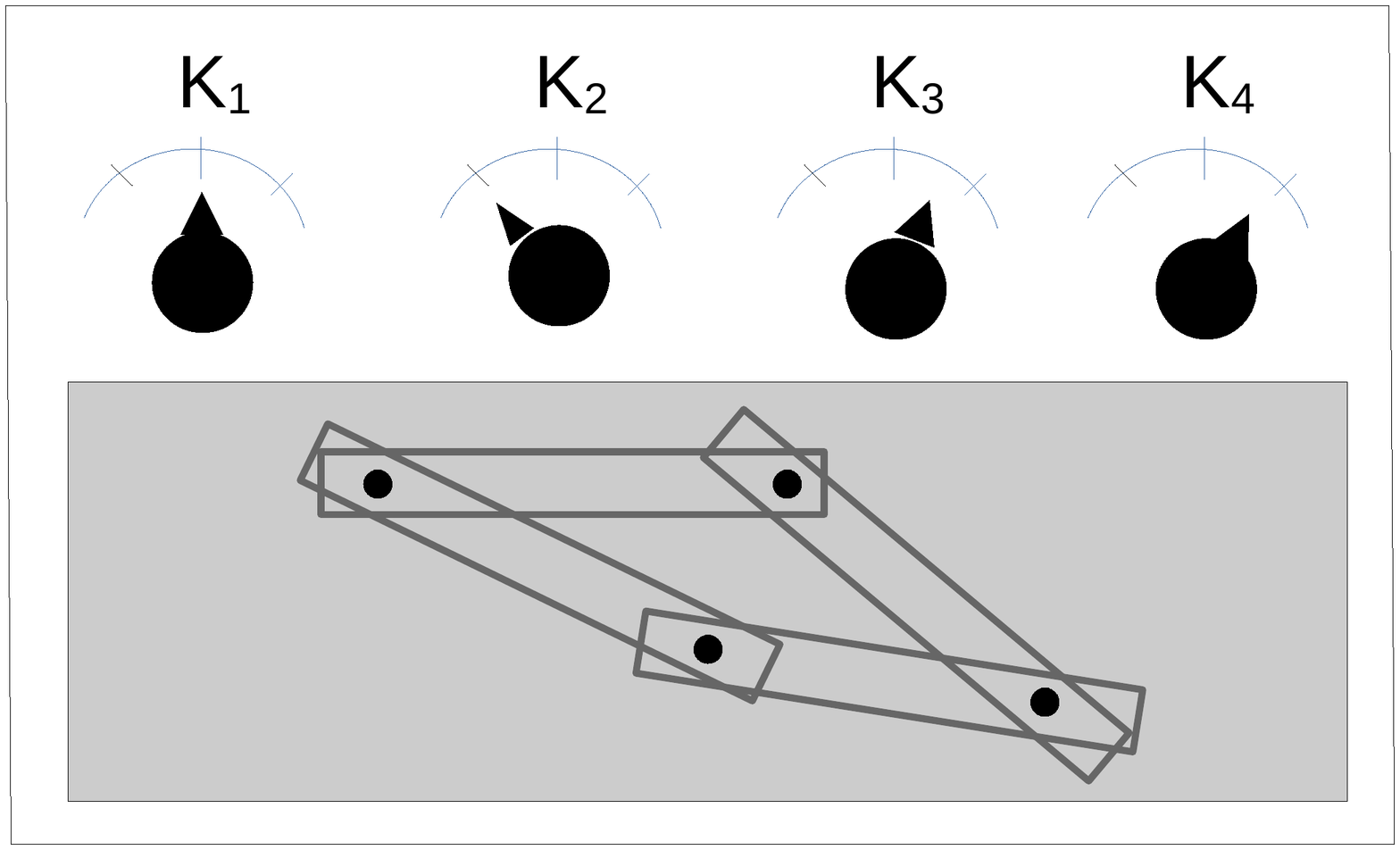}
\caption{\label{fig:apparatusmech} Apparatus whose front side contains $n$ measurement devices and $n$ knobs. The measuring device measure 
unknown quantities $X_1,\dots,X_n$. How do we `find out' (or `define'?) whether knob $j$ intervenes on $X_j$? The knobs in the toy model symbolize actions in the real world of which we want to define which variables they intervene on.} 
\end{figure}

Part of the fuzziness of causality in `high-level variables' in real-life applications can be captured by the following 
metaphoric toy example: 
to motivate Question \ref{q:vardef}, consider the apparatus show in figure \ref{fig:apparatusmech}. Instead of showing $n$ measuring instruments on its front side, it contains a window through which we see a mechanical device with a few arms and hinges, whose positions and angles turn out be controlled
 by the positions of the knobs. The angles and positions together satisfy geometric constraints by construction, they cannot be changed independently.\footnote{For causal semantics in physical dynamical systems see e.g. \cite{Bongers2018}.}  
 Accordingly, parameterizing the remaining $4$ degrees of freedoms via variables $X_1,\dots,X_4$  is ambiguous, e.g. horizontal and vertical position of one of the $4$ hinges and 2 angle (describing the system by more than $4$ variables would be over-parameterization, which results in constraints, as described 
 in Subsection \ref{subsec:coord}). 
 Now, the question whether or not turning knob $j$ can be seen as intervention of one particular mechanical degree of freedom $X_j$ depends on the parameterization.
 In this case, even hardware analysis does not necessarily clarify which degrees of freedom the knob intervenes on. 
The example shows also that there is an obvious `cheap' solution to the ill-definedness of interventions: we claim that the position of the knobs are the only variables
we are able to intervene on and only explore effects of those, while avoiding questions on causal relations between the {\it internal} variables 
(accordingly, one would deny talking about interactions {\it between genes} in gene expression experiments and consider 
changing experimental conditions (the `knobs') as the actual interventions). 
While this perspective sounds clean and circumvents hard conceptual problems of causality, it dismisses the idea of a causal understanding of the 
processes {\it inside the box}. 
\section{Ill-defined causal relations \label{sec:ill}} 

We  now try an incomplete taxonomy of reasons that render causality ill-defined, even after understanding the underlying processes. 
Using the apparatus in Figure \ref{fig:apparatus} as metaphor, we sometimes still don't understand causal relations even {\it after opening the box}. 
This is because causal relations between variables are not always like `wires that link devices'. 
We emphasize that some ill-definedness of causal relations in real life result from ill-definedness of the variable it refers to, thus leading us to 
Question 1. For instance, the question to what extent the air temperature outside today influences the value tomorrow: 
if `temperature today' is supposed to only refer to the temperature of a cubic decimeter around the temperature sensor, the impact is negligible. 
If, however, it refers to the mean temperature of the {\it whole region}, the impact is significant. While these temperatures largely coincide 
for passive observations, interventions destroy this equality and thus intervening on `the temperature'  is an ill-defined concept.
Let us now discuss a few reasons for ill-definedness of causal relations that appears even when the variables are well-defined. 

The purpose of providing the below incomplete list of reasons is to argue that discussions of causality often fail to
provide {\it enough context} to get a well-defined causal question. 
We will later see in what sense this context can be given
by specifying those elementary actions that we want to consider {\it interventions} on one variable.

\subsection{Coupling induced by 'coordinatization' of the world \label{subsec:coord}} 
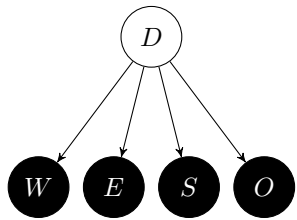
\begin{figure} 
\begin{center}
\begin{tikzpicture}
 \node[obs] at (-1.5,0) (W) {$W$} ;
 \node[obs] at (-0.5,0) (E) {$E$} ;
 \node[obs] at (0.5,0) (S) {$S$} ;
 \node[obs] at (1.5,0) (O) {$O$} ; 
 \node[var] at (0,2) (D) {$D$} edge[->] (W) edge[->] (E) edge[->] (O) edge[->] (S); 
\end{tikzpicture} 
 \end{center} 
\caption{\label{fig:hours} The hours spent with certain 'activities' like work, exercise, sleep, and others, are determined by a person's decision.} 
\end{figure}

For some person, let us define the variables $W,E,S,O$ describing the hours he/she spends 
for work, exercises, sleep, and others at some day. These variables satisfy the equation
\begin{equation}\label{eq:consist} 
W+E+S +O = 24,
\end{equation} 
a constraint which results in statistical dependences when observing the variables over several days. According to Reichenbach's principle of common cause \cite{Reichenbach1956}, each dependence is due to some {\it causal} relation, either the variables influence each other or they are influenced by common causes.
 Since one may be tempted to explain the statistical dependence by a common cause, let us introduce a hidden variables {\it the person's decision} influencing all $4$ variables,
as shown in Figure \ref{fig:hours}. 
However, this DAG suggests that independent interventions were possible. Obviously, the intervention of setting all
$4$ variables to $7$ hours is prohibited by \eqref{eq:consist}. Likewise, it is not possible to intervene on $W$ only 
without affecting the other variables. 
While these remarks seem obvious, they raise serious problems for defining downstream impact of the above variables, since it is impossible to isolate, for instance, the health impact of increasing $W$ 
from the health impact of the implied reduction of $E,S$, or $O$. 
Constraints that prohibit variable combinations a priori (by definition), rather than being a result of a mechanism (which could be replaced by others), are not part of the usual causal framework. The fact that independent changes of the variables are impossible also entails the problem of interventions being an ill-defined concept:
what qualifies a change of life-style which increased $E$ and decreased $W$ as {\it an intervention on} $E$? Is it a question of the intention?
If the intention to do more exercises entailed the reduction of work we tend to talk about an intervention on $E$, but given that people 
do not always understand even their own motivation behind their actions, it seems problematic that the
definition of an intervention then depends on these speculations.

\subsection{Dynamical systems}  
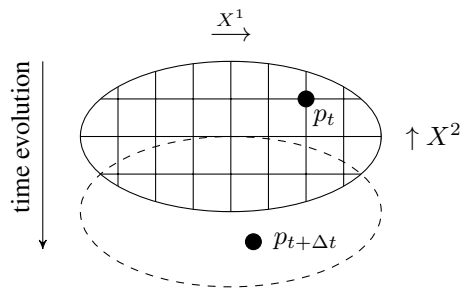
\begin{figure}
\begin{center}
\begin{tikzpicture}
\draw[<-] (-2.5,-0.5) -- (-2.5,2);
\node[rotate=90] at (-2.8,1) {time evolution};
\draw[dashed] (0,0) ellipse (2cm and 1cm);
\node at (0,2.5) {$\overset {X^1} \longrightarrow $};
\node at (2.7,1) {$\uparrow X^2$};
\node at (1.25,1.25) {$p_t$}; 
\filldraw (1,1.5) circle (3pt);
\node at (1,-0.4) {$p_{t+\Delta t}$}; 
\filldraw (0.3,-0.4) circle (3pt);
\draw[clip] (0,1) ellipse (2cm and 1cm);
\draw[step=0.5] (-5,-5) grid (5,5);
\end{tikzpicture}
\end{center} 
\caption{\label{fig:coord} A model of the world where the current state is a point in a topological space, that moves 
from point $p_t$ to $p_{t+\Delta t}$.} 
\end{figure}

Constraints  on variables like the ones above result naturally if we don't think of the state of the world as a priori given {\it in terms of variables}. Let us, instead, consider a model where the admissible states of the world are given as points in a topological space (in analogy to the {\it phase space} of a classical dynamical system \cite{Goldstein2002}),
in which variables arise from introducing coordinates, as visualized in Figure \ref{fig:coord}. 
Here, the ellipse defines the admissible states of the world and we have introduced a coordinate system
whose coordinates define the two random variables $X^1$ and $X^2$. 
The states of our toy world are formally given by the set of all 
those pairs $(x^1,x^2)$ that belong to the ellipse. 
We assume that the dynamics of our toy world is given by a topological dynamics
of the ellipse, that is, a continuous map that maps each state $p_t\in \R^2$ to its time evolved state $p_{t+\Delta t}$. 
The fact that the state cannot leave the ellipse entails a relation between the variables $X^1$ and $X^2$ 
of which it is unclear whether we should consider it a {\it causal} relation. We will therefore discuss how to define the
 notion of
'interventions' on $X^1$ and $X^2$. 

It is natural to interpret the intervention 'set $X^1$ to $\tilde{x}^1$ as an operation that maps a state $(x^1,x^2)$ to $(\tilde{x}^1,x^2)$. This is, however, only possible if 
$(\tilde{x}^1,x^2)$ is still a point in the ellipse. Otherwise, one is forced to change $x^2$ to some other value $\tilde{x}^2$ 
to still end up in an admissible state of the world. Could we then say that $X^2$ changed {\it because} 
$X^1$ changed? This interpretation may be valid if someone's intention was to change $X^1$, who is then forced to also change $X^2$ by the existing constraints. However, without this knowledge about what has been the intention of the actor, 
we just observe that both variables have been changed. Interpreting the action as an intervention on $X^1$ becomes questionable. Our simple model of the world already suggests two different notions of causality:

\begin{enumerate}
\item {\bf Causality between time-like measurements}  Here we consider the variables $X^1_t,X^2_t$ 
as causes of $X^1_{t+\Delta t},X^2_{t+\Delta t}$. 
\item {\bf Causality without clear time-separation}  Here, causal relations between the coordinates
appear as a phenomenon emerging from consistency conditions that define the admissible states of the world. 
\end{enumerate} 
Note that a more sophisticated version of constraints for dynamical systems can arise from equilibrium conditions \cite{DGL} since
the set of active constraints are active only for some interventions, while they get inactive for others \cite{Blom2019}, which
motivates so-called 
``Causal Constraints Models (CCMs)''. 
The fact that $X^1_t$ and $X^2_t$ refer to the same points in time, suggests to attribute their observed statistical dependences 
(which result from most distributions on the ellipse) to their common history and draw the causal DAG
\[
X^1 \leftarrow H \rightarrow X^2,
\]
where $H$ encodes the relevant feature of the state $(X^1_{t-1},X^2_{t-1})$.  
However, this DAG suggests the existence of independent interventions on $X^1_t$ and $X^2_t$,  although the constraints 
given by the ellipse need to be respected by any action (similar to our remarks on Figure \ref{fig:hours}).  This fact would rather be covered by a
 causal chain graph containing an undirected link $X^1_t - X^2_t$, as discussed in \cite{Lauritzen2002} for mutual interaction in equilibrium states. 
If we think of a force in the direction of the $x^1$-axis and observe that the point moves also in $x^2$-direction 
once it reaches the boundary, we would certainly consider $X^1$ the cause and $X^2$ the effect. 
However, once the boundary is reached it is no longer visible that its a force in $x^1$-direction that drives the curved motion. Accordingly, the causal direction becomes opaque. 

The second case is more interesting for this paper since it refers to a notion of causality that
is less understood. At the same time, it challenges the interpretation of causal directions as a concept
that is necessarily coupled to time order. Instead, this kind of {\it phenomenological} causality can also emerge from 
relations that are not given by the standard physical view on causality dealing with {\it signals} that propagate trough
{\it space-time} from a sender to a receiver.\footnote{which prohibits instantaneous influence between remote objects since no signal can propagate faster than light \cite{Einstein1920}.} (such a physically local concept of causality admits defining interventions on a variable $X_j$ as actions for which a signal from the actor
reaches the location of $X_j$). 
There is no analogous approach for phenomenological causality, since it may be too abstract to be directly materialized in space-time. For these reasons,
Question \ref{q:intervention} should be particularly raised for causality in domains outside physics when referring to sufficiently abstract variables.
However, the variables  $X^1$ and $X^2$ referring to different coordinate of the same system 
can also have an abstract causal relation.\footnote{Note that \cite{Bongers2022} also discusses interventions that change some coordinates
 in dynamical systems, and asks whether position and momentum of a physical particle allow for separate interventions.}

 \subsection{Undetectable confounding: distinction between a cause and its witness} 
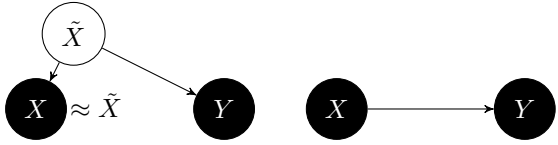
\begin{figure} 
\begin{center}
\begin{tikzpicture}
 \node[obs] at (-1.5,0) (X) {$X$} ;
 \node[obs] at (1,0) (Y) {$Y$} ; 
 \node[var] at (-1,1) (tX) {$\tilde{X}$} edge[->] (X) edge[->] (Y); 
 \node[] at (-0.7,0.05) {$\approx \tilde{X}$};
\end{tikzpicture} 
\hspace{0.5cm} 
\begin{tikzpicture}
 \node[obs] at (1,0) (Y) {$Y$} ; 
 \node[obs] at (-1.5,0) (X) {$X$} edge[->] (Y); 
\end{tikzpicture} 

 \end{center} 
\caption{\label{fig:noisyM} If $X$ is an arbitrarily perfect copy of the confounder $\tilde{X}$ (e.g. if $X$ is a reasonably good measurement of the true quantity $\tilde{X}$), distinction between the left and the right scenario gets arbitrarily hard.} 
\end{figure}

 To distinguish between the scenarios $X\rightarrow Y$ and $X  \leftarrow \tilde{X}    \rightarrow Y$ where $\tilde{X}$ is latent, 
is one of the most relevant and challenging problems. Methods have been proposed that address this task from passively observing $P(X,Y)$ 
subject to strong assumptions, e.g., \cite{HoyerLatent08,UAI_CAN}  or in scenarios where $X,Y$ are embedded in a larger network of variables, e.g., the Fast Causal Inference Algorithm \cite{Spirtes1993}, or instrumental variable based techniques \cite{Bowden} and related approaches \cite{Atalanti}. 
Certainly the task gets arbitrarily hard when the effect of $\tilde{X}$ on $X$ gets less and less noisy, see Figure \ref{fig:noisyM}, left.  
In the limiting case where $X$ is an exact copy of $\tilde{X}$, no algorithm
can ever tell the difference between the two scenarios in Figure \ref{fig:noisyM}  from passive observations. Note that this scenario is quite common when $\tilde{X}$ is some physical quantity and $X$ the value of $\tilde{X}$ shown by a precise measurement device. In this case, we would usually not even verbally distinguish between $X$ and $\tilde{X}$ although
it certainly matters whether an intervention acts on $X$ or $\tilde{X}$. The distinction gets only relevant when we act on the system, and even then, only if actions are available that decouple
$X$ from $\tilde{X}$. Accordingly, it is the set of available actions that defines causality, which is the crucial idea in Section \ref{sec:element}. 

A similar ill-definedness occurs in scene understanding in computer vision: 
Imagine a picture of a dog snapping for a sausage. Most people would agree that the sausage is the {\it cause} for the presence of the dog (unless the scene is taken from a location where the dog commonly stays anyway, e.g., its doghouse).
However, the presence of the sausage {\it on the image} is certainly not the cause for the presence of the dog on the image -- changing the respective pixels
in the image will not affect the pixels of the dog. Similar to the measurement device, we can only meaningfully talk about causal relations if we 
do not distinguish between the cause `presence of the sausage'  and its `witness', e.g., its occurrence on the image. Ignoring 
actions that decouple the presence of an object in the real scene from its appearance on the image (retouching), we may also  consider
the presence of the sausage on the image the cause of the presence of the dog on the image and call this causal relation `phenomenological'.

 \subsection{Coarse-grained variables \label{subsec:coarse}}
 Whenever coarse-grained (`macrosocopic') variables are defined, for instance, by averaging over `microscopic' variables,
 interventions on the former are no longer well-defined since different changes on the micro-level amount to the same change of the macro-variable.
 Thus, causal relations  between macro-variable may be ill-defined.  
Accordingly,  \cite{Rubensteinetal17} state a consistency condition for coarse-grainings of causal structures according to which interventions on 
micro-variables that result on the same intervention on the macro-variables, also entail the same downstream impact on other 
macro-variables. \cite{Beckers2019} argue that different strength of consistency conditions are needed for different levels of abstraction.  
We believe that discussions on causal relations in real life often refers to macro-variables for which the impact of interventions does highly depend  
on how the intervention is implemented, and it is thus hard to identify {\it any} valid consistency condition, however weak it may be. 
Let us consider the following toy model. Given the variables $X_1,X_2,Y_1,Y_2$, with the causal DAG shown in
\eqref{eq:coarse}, where $X_1$ influences $Y_1$ and $Y_2$ influences $X_2$:   
\begin{eqnarray}\label{eq:coarse} 
\bar{X}  \left\{ \quad \begin{array}{ccc} X_1 & \longrightarrow  & Y_1 \\
X_2 &\longleftarrow & Y_2 \end{array} \quad \right\}  \bar{Y}.
\end{eqnarray}  
We assume the FCMs 
\begin{eqnarray*} 
Y_1 &=& X_1\\
X_2 &=& Y_2,
 \end{eqnarray*} 
and define the macro-variables $\bar{X} := (X_1+X_2)/2$ and $\bar{Y}:= (Y_1 +Y_2)/2$, whose causal relations we want to discuss. 
Obviously, neither an intervention on $\bar{X}$ nor on $\bar{Y}$ is well-defined. Increasing $\bar{X}$ by $\Delta$ can be done by
 adding any vector of the form $(\Delta + c, -c)$  with $c\in \R$, and likewise for intervention on  $\bar{Y}$.
 Someone who changes $\bar{X}$ and $\bar{Y}$ by changing $X_1$ and $Y_1$, respectively, will claim that $\bar{X}$ influences
 $\bar{Y}$, while someone changing the macro-variables by acting on $X_2$ and $Y_2$ only considers $\bar{Y}$ the cause of $\bar{X}$.
 We conclude that not only the quantitative effect but even the causal direction is not a property of the system alone, but a result of
 which actions are available.

\subsection{Diversity of non-equivalent interventions} 

In the framework of FCMs and graphical  causal models, the impact of interventions (e.g. point interventions $do(X_j=x_j)$) 
does not depend on {\it how} this intervention has been performed. Here, the word `how' is meant in the sense of {\it which
mechanism has been used to change $X_j$}. 
The description of  the mechanism implementing the intervention is not part of the description, also because the framework implies that different ways 
of setting $X_j$ to $x_j$ have the same impact. In real world problems, however, we talk about impact of one variable on another one without specifying 
the model we refer to, which renders impact of interventions ill-defined. 
Let us elaborate on this in the context of stochastic processes. Let $(X_t)_t$ be a time series of the electricity consumption of a household, where each value is the integral over one hour.  A simple action to reduce $X_t$ at some $t$ could be to convince the resident 
not to start his/her dish washer at this point in time. Whether or not this action causes a change of $X_s$ at some later time $s>t$ depends on whether 
the resident decides to clean the dishes by hand or to just delay the start of the machine. 
Likewise, the impact of changing the traffic of some road depends on {\it how} it is performed. In a scenario where the road is made less attractive by a strong speed limit, drivers may take a different route and thus increase the traffic of other roads. Reducing the traffic by offering additional public transport would not have the same effect. Again, it is the nature of the action that defines the causal impact, instead of some causal truth that holds without this additional specification.

\subsection{Abstract causal mechanisms} 
While the previous subsection described difficulties with the concept of causality that already arise  in clearly defined physical systems, we 
now discuss an example where the causal mechanisms lie in a more abstract domain.  
Examples of causal relations for which we believe the simple mechanistic view of causality to be problematic, are widespread
in the literature. \cite{Schoelkopf2021}, for instance, describe an scenario of online shopping where a laptop 
is recommended to a customer who orders a laptop rucksack. \cite{Schoelkopf2021} argue that this would be odd because
the customer probably has a laptop already. Further, they add the causal interpretation that buying the laptop 
is the {\it cause} of buying the laptop rucksack. 
We do agree to the causal order but do not believe the {\it time order} of 
the purchases to be the right argument. For someone who buys laptop and laptop rucksack in different shops it can be reasonable to buy the rucksack first in order to
safely carry the laptop home (given that he/she already decided on the size of the laptop). We believe, instead, that 
the recommendation is odd because the {\it decision} to purchase the laptop is the cause of the {\it decision} to buy the rucksack. 
Whether or not this necessarily implies that the decision has been made earlier, is a difficult question of brain research. If they were made in two well-localized, slightly different 
positions in the brain, one could, again, argue that causal influence can only be propagated via a physical signal (of finite speed).
We do not want to further elaborate on this question. The remarks were only intended to show that causal problems in everyday business processes refer to rather abstract notions of causality -- for instance, to causality between {\it mental} states. It seems that particularly 
in these domains, causality seems to be a particularly context-dependent concept. 

Economic variables often show a several of the above aspects of ill-definedness coming from aggregation or psychological factors or both.
 For example, the price of a particular good is not only understood as the price at which one firm sells that good (unless we are in a monopoly market) but instead as an aggregation of prices. Likewise, Consumer Confidence Indices (CCI) are an aggregation of the beliefs of individual agents. Economic indices tend to be more abstract than their name might suggest.

\section{\label{sec:im} Complexity aspect of Independence of Mechanisms (IM)} 
The idea that the conditionals $p(x_j|pa_j)$ in \eqref{eq:fac} correspond to `independent' mechanisms of the world, has a long tradition in the causality community, see
e.g. \cite{causality_book}, section 2.2, for different aspects of `independence' and their history. \cite{Algorithmic,LemeireJ2012} conclude that the different conditionals contain no algorithmic information about each other, \cite{anticausal} conclude that they change independently across environments and describe implications for transfer learning scenarios.  The `sparse mechanism shift hypothesis' \cite{Schoelkopf2021} assumes that changing the setup of an experiment often results in changes of $p(x_j|pa_j)$ for a small number of nodes $X_j$.

Here we want to discuss this independent change from a slightly different perspective, namely from the one
of {\it elementary} versus {\it complex} actions. To this end, we restrict the attention to a bivariate causal relation $X\to Y$. 
According to the interpretation of IM in \cite{anticausal}, the causal structure entails that $P(X)$ and $P(Y|X)$ change {\it independently} across environments.
More explicitly, knowing that $P(X)$ changed to $P'(X)$ between training and test data, does not provide any information on how $P(Y|X)$ changed. In absence 
of any further evidence, it will thus often be reasonable to assume that $P(Y|X)$ remained the same (which is the so-called covariate shift scenario \cite{Masashi2012}).
Likewise, it can also be the case that $P(Y|X)$ changed to $P'(Y|X)$ while $P(X)$ remained the same. However, the scenario that only $P(Y)$ changed and $P(X|Y)$ remained the same or vice versa, is rather unlikely. The reason is that this required contrived {\it tuning} of the changes of the mechanisms of 
$P(X)$ and $P(Y|X)$. Let us illustrate this idea for a simple example. 

\begin{Example}[ball track]\label{ex:ball_track} 
Figure \ref{fig:ball_track} is an abstraction of a real experiment (which is one of the cause-effect pairs in \cite{cepairs}) with a ball track.  
A child puts the ball on the track at some position $X$, where it accelerates and reaches a point where its 
velocity $Y$ is measured by two light barriers. 
\begin{figure} 
\includegraphics[width=0.5\textwidth]{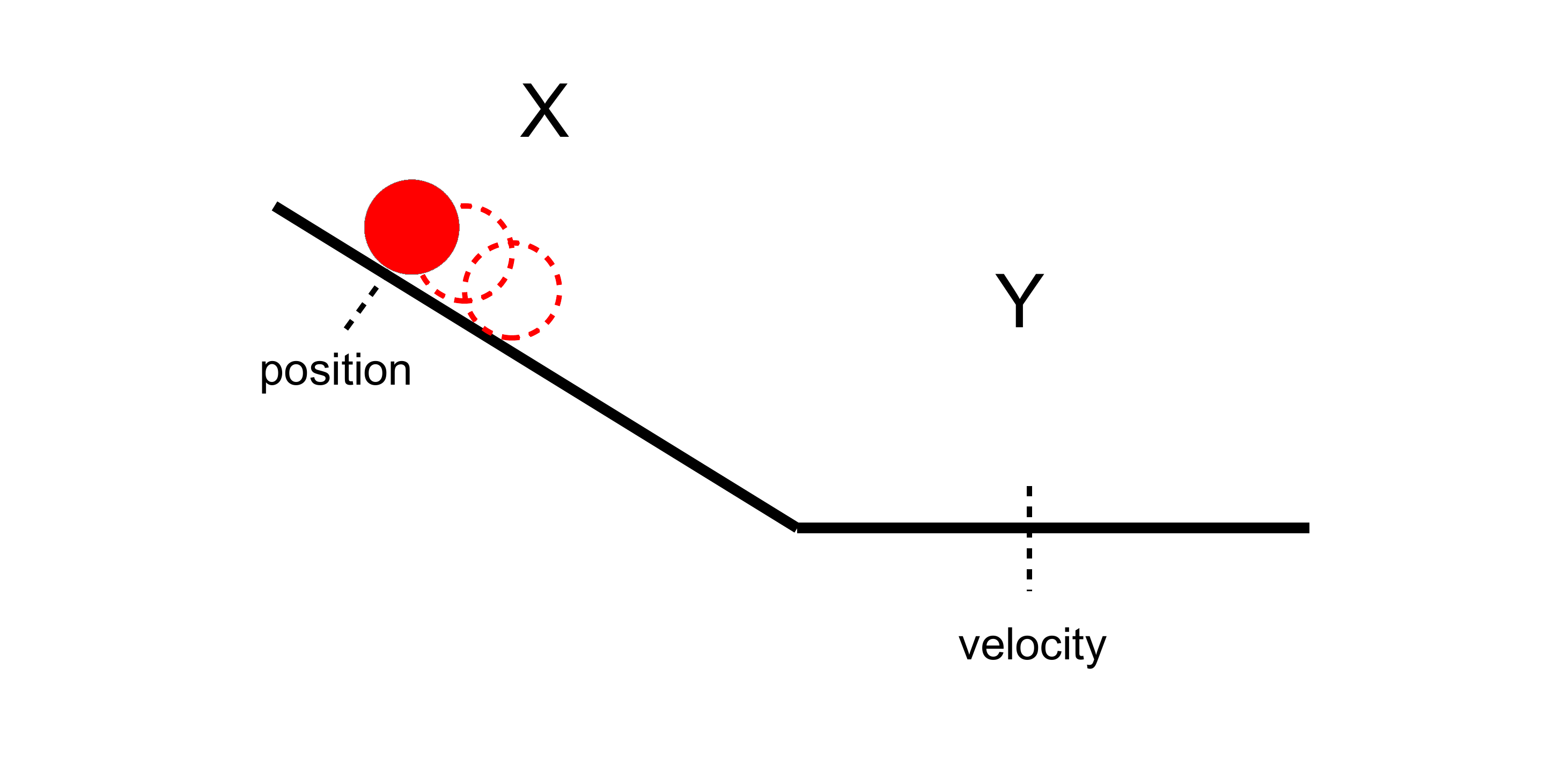}
\caption{\label{fig:ball_track} Cause-effect pair with ball-track, taken from \cite{cepairs}: The cause $X$ is he start position of the ball along the inclined plane and the effect $Y$ is the speed at which the ball passes the light barriers at the horizontal track. The example illustrates that $P(X)$ and $P(Y|X)$ correspond to independent mechanisms.} 
\end{figure} 
One can easily think of a scenario where $P(X)$ changes without affecting $P(Y|X)$ 
from datasets to the other one: an older child will tend to choose positions $X$ that are higher. 
On the other hand, changing $P(Y|X)$ without affecting $P(X)$ can be done, for instance, by mounting the light barriers at a different position and thus measuring velocity at a later point where the ball already lost some speed. It requires, however, contrived actions to change $P(Y)$ without changing $P(X|Y)$. This would involve
both changes of the child's behaviour {\it and} changes at the speed measuring unit. 
\end{Example}
Example \ref{ex:ball_track} shows a complexity aspect of IM that we want to build on throughout the paper:
changing $P(X)$ or $P(Y|X)$ without affecting the other is easy and requires only {\it one} action. In contrast, changing $P(Y)$ or $P(X|Y)$ without changing
the other one of these two objects is difficult for two reasons: first, it requires changes of both, the distribution $P(X)$  of start positions and the conditional 
$P(Y|X)$ via shifts of the speed measurement.
Second, these two actions need to be tuned against each other. After all, those actions on $P(X)$ and $P(Y|X)$  that are easy to implement
(e.g. replace the child, shift the mounting of the light barrier) will probably not match together in a way that affects {\it only} $P(Y)$ but not $P(X|Y)$. 
In general, if we assume that not all operations on $P({\rm Cause})$ and $P({\rm Effect}|{\rm  Cause})$ are elementary, it may thus take 
even a {\it large number} of operations to change only $P({\rm Effect})$ without affecting $P({\rm Cause}|{\rm Effect})$.    
Accordingly, for causal DAGs with $n$ nodes, we assume that all elementary operations change at most one conditional $P(X_j|PA_j)$,
 but we do not assume that any change of a single conditional is elementary. Further, we do not even assume that {\it any arbitrary} change of 
$P(X_j|PA_j)$ can be achieved by concatenations of elementary actions. 

To relate this view to known perspectives causal counterfactuals, note that Lewis \cite{Lewis1979} defined the impact of an event $E$ via a hypothetical world that is most similar to the true one except for the fact that $E$ did not happen, as opposed to a world in which  $E$ happened, but also several subsequent actions were taken so that the world gets back to the path it would have followed without $E$. In the spirit of our paper, we could think of $E$ as generated by one elementary action and read Lewis' view as the statement that after one elementary action
the world is closer in Lewis' sense to the original one  than after several interventions that undo the downstream impact of the first one.

\section{\label{sec:element} Defining causal directions via elementary actions} 
Here we describe the main idea of the paper which uses the notion of `elementary action' as first principle, and then discuss quite diverse toy examples.
Some of them are directly motivated by practically relevant real-life applications, but we also discuss strongly hypothetical scenarios, only constructed with the purpose of challenging our intuition on causality. 

\subsection{The bivariate case} 
To avoid the above circularity of defining interventions in a way that relies on the concept of causality and the other way round, we suggest the following approach: 
\begin{Idea}[phenomenological cause-effect pair]\label{idea:Acausality} 
Let $X,Y$ be two variables describing properties of some system $S$ and  $\cA$ be a set of elementary actions on $S$.
We say that $X$ causes $Y$ whenever $\cA$ contains only the following two types of actions:\\
\noindent
$\cA_1:$ actions that change $X$, but preserve the relation between $X$ and $Y$\\
\noindent
$\cA_2:$  actions that preserve $X$, but change the relation between $X$ and $Y$. 
\end{Idea}  
Since Idea \ref{idea:Acausality}  is quite informal, it leaves some room for different interpretations. We will work with two different
ways of spelling it out:

\begin{Definition}[statistical phenomenological causality] \label{def:phcestat} 
Let $X,Y$ be two variables describing properties of some system $S$ and  $\cA$ be a set of elementary actions on $S$.
We say that $X$ causes $Y$ whenever $\cA$ contains only the following two types of actions:\\
$\cA_1:$ actions that change $P(X)$, but preserve $P(Y|X)$ and \\
$\cA_2:$  actions that preserve $P(X)$, but change $P(Y|X)$.
\end{Definition} 

\begin{Definition}[Unit level  phenomenological causality]\label{def:pheceunit} 
We say that $X$ causes $Y$ whenever $\cA$ contains only the following two types of actions:\\
$\cA_1:$ a set of actions (containing the identity) such that every pair $(x',y')$ obtained from the observed pair $(x,y)$ by an action in $\cA_1$ 
satisfies the same law $y'= m(x')$ for 
some (non constant) function $m$.\\ 
$\cA_2:$ actions that keep $x$.  
\end{Definition} 
Note that $m$ in Definition \ref{def:pheceunit} holds for all actions in $\cA_1$, but different functions $m$ hold for different statistical units.   
If we think of $X$ and $Y$ as related by the FCM $Y= f(X,N)$ we should think of $m$ as the map $f(.,n)$ with fixed noise value $n$. 
The statement that actions in $\cA_1$ do not change the mapping from $X$ and $Y$ thus refers to the counterfactual knowledge encoded by the FCM, which we
assume to be given from domain knowledge about the system\footnote{We think this is justified because 
the scope of this paper is to discuss how to define causality, not how to infer it.}.
Further note that changing the map $m$  can be done by either changing $f$ or $n$.
Although the condition for $\cA_1$ is asymmetric with respect to swapping $X$ and $Y$ since $m$ maps from $X$ to $Y$,
this asymmetry does not necessarily imply the mapping $m$ to be {\it causal}. 
Assume, for instance, $X$ and $Y$ are related by the FCM $Y= X+N$. Then, an observed pair $(x,y)$ for which $N=3$ will obey the rule
$y= x+ 3$ and $y' = x' +3$ for all pairs generated by actions in $\cA_1$. However, all these pairs will also obey the rule $x' = y' -3$, and thus there exists also
a map $\tilde{m}$ from $Y$ to $X$.  In our examples below the crucial asymmetry between cause and effect will not be induced by the existence of $m$, but by the existence of actions $\cA_2$, which only act on the effect. In other words, interventions on the cause to do not reveal the asymmetry because they change cause and effect, while actions on the effect only change the effect. 

As an aside, note that for the scenario where $X$ and $Y$ are only connected by a confounder we would postulate actions that affect only $X$ and those that affect only $Y$.

The idea of identifying causal structure by observing which conditionals in \eqref{eq:fac} change independently across datasets can already be found in the literature, e.g., \cite{Zhang2017CausalDF} and references therein.  In the same spirit, Definitions \ref{def:phcestat} and  \ref{def:pheceunit} raise the question whether 
they define the causal direction uniquely. This is easier to discuss for Definition  \ref{def:phcestat}. Generically, changes of $P(X)$ results in simultaneous changes of both $P(Y)$ and $P(X|Y)$, which thus ensures that the available actions of class $\cA_1$ neither fall into the category corresponding to $\cA_1$  nor $\cA_2$ for the backwards direction from $Y\to X$. 
The following simple result shows a genericity assumption for which this can be proven: 

\begin{Proposition}[identifiability via changes]  
Let $X$ and $Y$ be finite with $|\cX| = |\cY|$ and the square matrix $p(x,y)_{x,y} $ have full rank with $p(x,y)$  strictly positive. 
Then changing $p(x)$ 
changes $p(y)$ and $p(x|y)$. 
\end{Proposition}
\noindent
\begin{Proof}  
Define $\tilde{p}(x,y):=\tilde{p}(x) p(y|x)$. 
Assume 
$
\tilde{p}(x|y) = p(x|y).
$ 
Hence, 
\[
\tilde{p}(x) p(y|x) \tilde{p}^{-1} (y) =   p(x) p(y|x) p^{-1} (y),  
\]
which is equivalent to 
$
\tilde{p}(x)  p (y) =   p(x) \tilde{p} (y).
$
Summing over $y$ yields $\tilde{p}(x) =p(x)$, hence $p(x)$ did not change.
We conclude that changing $p(x)$ changes $p(x|y)$. That changing $p(x)$ also changes $p(y)$ follows from the full rank assumption.
\end{Proof}

\paragraph{Abstract toy example} 
We now describe an example for unit level phenomenological causality according to Definition \ref{def:pheceunit} where the causal direction
is a priori undefined, but may be defined after specifying the set of actions, if one is willing to follow our approach. By purpose, we 
have chosen an example whose causal interpretation seems a bit artificial. 
\begin{Example}[urn model] \label{ex:urn} 
Assume we are given an urn containing blue and red balls, as well as a reservoir containing also blue and red balls. The game allows four basic operations: 
 ($A_1^+$) replacing a red ball in the urn with a blue one, ($A_1^-$) 
 replacing a blue ball with a red one,
 ($A_2^+$) adding a red ball to the urn, and 
($A_2^-$) removing a red ball from the urn (and adding it to the reservoir),
see Figure \ref{fig:urns}.
\begin{figure} 
\includegraphics[width=0.24\textwidth]{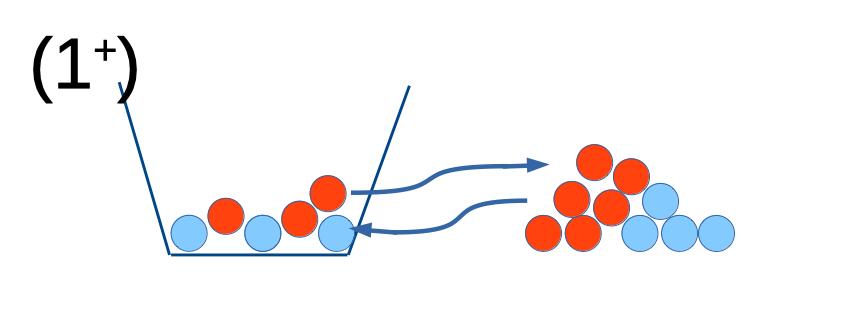}
\includegraphics[width=0.24\textwidth]{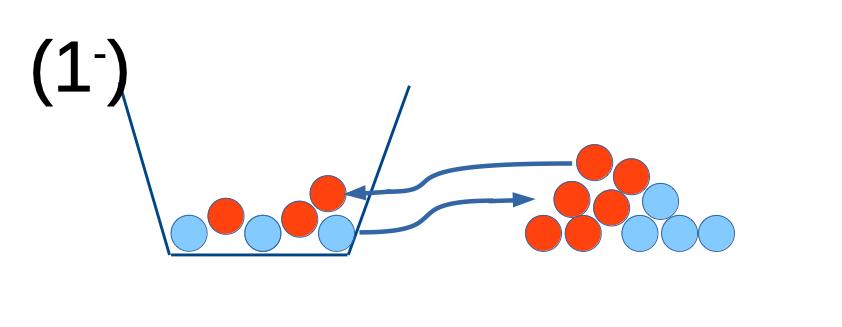}\\
\includegraphics[width=0.24\textwidth]{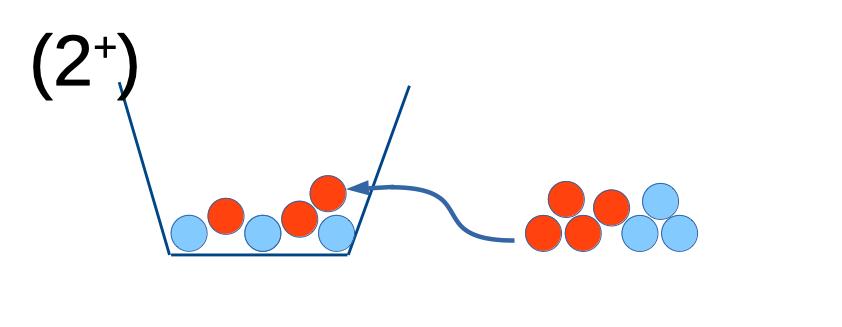}
\includegraphics[width=0.24\textwidth]{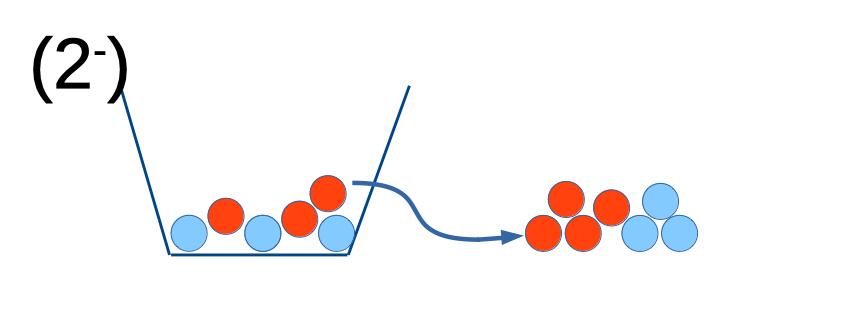}
\caption{\label{fig:urns} Urn model in Example \ref{ex:urn} with $4$ different operations.} 
\end{figure} 

Define the random variables $K_b$ and $K_r$, describing the number of blue and red balls in the urn, respectively.
 Obviously, the 4 different operations correspond to the following 
changes of $K_b,K_r$:
\begin{eqnarray*}
(A^+_1) \quad K_b &\to& K_b+1; \quad K_r \to K_r-1\\
(A^-_1) \quad K_b &\to& K_b-1; \quad K_r \to K_r+1\\ 
(A^+_2) \quad K_r &\to & K_r+1 \\
(A^-_2) \quad K_r &\to& K_r-1. \\ 
\end{eqnarray*}
Note that action $A_2^+$ is always possible, but the other three operations are only possible if the quantity to be reduced is greater than zero.
According to Definition \ref{def:pheceunit}, 
we have the causal relation $K_b\rightarrow K_r$ because 
the actions $A_1^\pm$ belong to the category $\cA_1$ since they preserve the relation $K_b = c - K_r$ 
for some state-dependent constant $c$. Further, $A_2^\pm$ belong to the category $\cA_2$. 
We also observe  that changing $K_b$ without changing $K_r$ requires the concatenation of
two operations at least: For instance, add a red ball, and then `convert' it into a blue one. 
\end{Example}
The example may be considered as representing a chemical process where
molecule of type $K_r$ can be converted into molecule of type $K_b$ and vice versa. 
Then, the `red' molecules are the resource for a reaction that converts `red' into `blue'. 
Therefore,  one may be surprised 
that not the resource $K_r$, but the product $K_b$ of the reaction, is the cause. 

We now rephrase Example \ref{ex:urn} into an example for the statistical version in Definition \ref{def:phcestat}. To this end, we consider a random experiment for which the system is initially in the state $K_r=k_r$ and $K_b=k_b$  with $k_r,k_b\gg 0$. Then, in each round we flip a coin for each of the $4$ actions to decide 
whether they are applied or not.  After $\ell < k_r,k_b$ many rounds, let $N_1$ denote the number of times $A_1^+$ minus the number of times $A^-_1$ has been applied. Likewise, $N_2$ counts the number of $A^+_2$ minus the number of $A^-_2$ actions.
We then obtain 
\begin{eqnarray} 
K_b &=& k_b + N_1 \label{eq:bica} \\
K_r &=& k_r - N_1 + N_2. \label{eq:rica}
\end{eqnarray} 
On easily checks that \eqref{eq:bica} and \eqref{eq:rica} are equivalent to 
\begin{eqnarray} 
K_b &=& k_b + N_1 \label{eq:bse} \\
K_r &=& - K_b + k_r + k_b + N_2, \label{eq:rse}.
\end{eqnarray} 
Since the actions are controlled by independent coin flips, we have  $N_1\independent N_2$.
Following our interpretation that $K_b$ causes $K_r$ and $A^\pm_1$ and $A^\pm_2$ are interventions on 
$K_r$ and $K_b$, respectively, we thus consider \eqref{eq:bse} and \eqref{eq:rse} as the corresponding FCM.
By controlling actions $A^\pm_1$ and $A_2^\pm$ via coins with different bias, we may change the distributions 
 $P(K_r|K_b)$ and $P(K_b)$ independently, and thus have an example of 
statistical phenomenological causality in Definition  \ref{def:phcestat}.

The following observation may seem paradoxical at first glance: The set $\cA_1=\{A_1^+,A^-_1\}$ is a priori symmetric with respect to swapping the roles
of blue and red. The justification for calling it interventions on $K_b$ is derived from properties of $\cA_2=\{A_2^+,A_2^-\}$.  
In other words, whether an action is considered
an intervention on a certain variable depends on the impact of other actions in the set of elementary actions. 
This context-dependence of the definition of interventions may be worrisome, but in scenarios where 
`hardware-analysis' does not reveal (or define) whether an action is an intervention on a particular variable, we do not see a  chance that circumvents 
this dependence on other actions. 
Given the abstractness of the underlying notion of causal direction, we are glad to observe that the well-known causal discovery method LinGAM 
\cite{Kano2003} 
would also infer $K_b \rightarrow K_r$ because \eqref{eq:bse} and \eqref{eq:rse} define a linear model with non-Gaussian additive noise. 
The crucial assumption inducing the statistical asymmetry is that actions $A^\pm_1$ are implemented independently of $A^\pm_2$, resulting in 
independent noise variables $N_1,N_2$. 

There is also another aspect of this example that shows the abstractness of the causal interpretation of the above scenario. The fact that actions 
in $\cA_1$ preserve the total number of balls has been interpreted as structural equation \eqref{eq:rse} generating $K_r$ from $K_b$. Using the function
$m$ from Definition \ref{def:pheceunit}, this structural equation reads $K_r= m(K_b)$ with $m(K_b) = - K_b + k_b$. 
Since operations in $\cA_2$ change the total number of balls, they change  $m$ to $m'$ by changing $k_b$. Hence, actions in $\cA_2$ change the `mechanism'
relating $K_b$ and $K_r$. In a mechanistic interpretation of causality, one would expect changes of a mechanism a change of a kind of machine where
the input-output behaviour is changed. As abstract as the `mechanism' from $K_b$ to $K_r$ is, as abstract is its change.

 \paragraph{Context-dependent causal directions}  
Here we describe a system for which causal directions swap when the system moves from one regime to another one.
Although the following example is hypothetical, we encourage the reader to think of similar example in realistic business processes.
\begin{Example}[food consumption of rabbits]\label{ex:rabbits} 
Given a hutch with $n$ rabbits where we define two variables:
\begin{tabular}{ll}
$X$:& total amount of food consumed by all rabbits at\\ &one day\\
$Y$:& food per rabbit consumed at one day. 
\end{tabular} 
By definition, we have $Y = X/n$.
We allow the following three types of actions: \\
$\cA_r$:  change the number $n$ of rabbits\\
$\cA_f$: change the amount of food provided\\
$\cA_a$: give an appetizer to the rabbits.\\  
We then consider two complementary scenarios, see Figure \ref{fig:rabbits}:
\begin{figure} 
\includegraphics[width=0.24\textwidth]{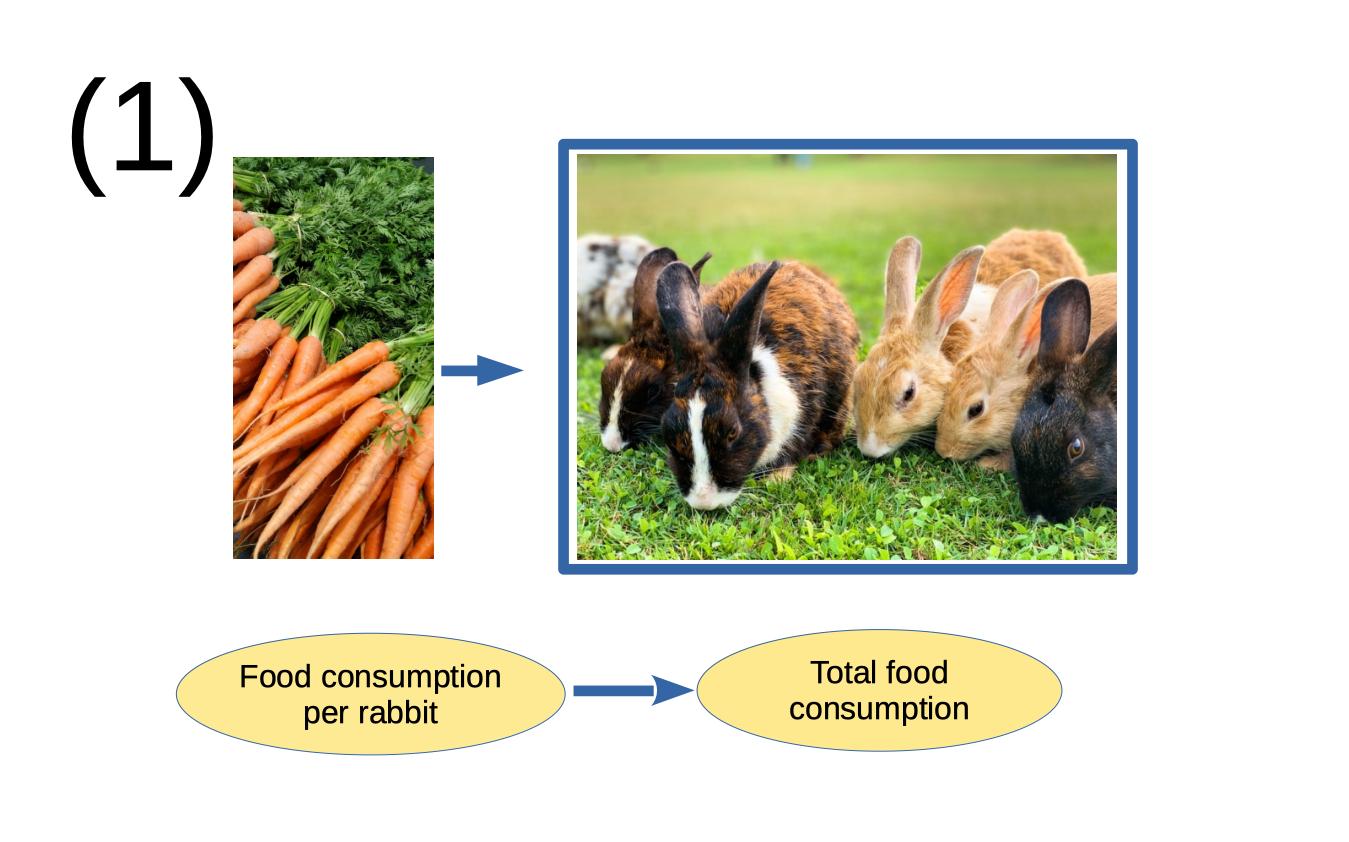}
\includegraphics[width=0.24\textwidth]{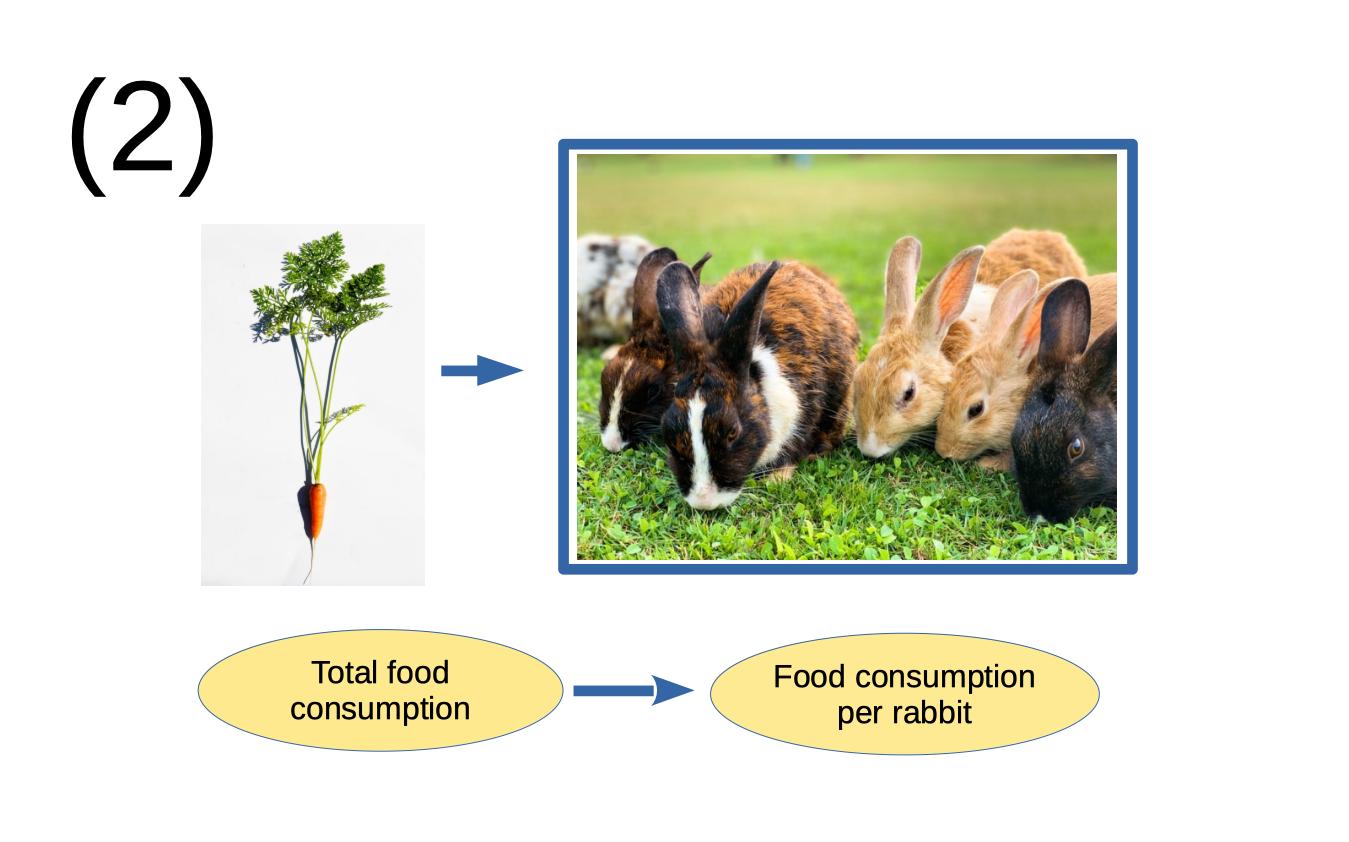}
\caption{\label{fig:rabbits} In scenario 1, the food consumption per rabbit is causing the total food consumption because
changing the number of rabbits only changes the latter if there is enough food for each rabbit. In scenario 2 with food shortage, changing the number of rabbits does not affect the total food consumption, it only changes the food consumption per rabbit. Therefore, the total food consumption is the cause. Images by 
Tom Paolini (carrots),  michealcopley03 (single carrot),  
Aswathy (rabits) with
 unsplash license. 
} 
\end{figure} 

\vspace{0.1cm}
\noindent
{\bf Scenario 1: there is enough food for each rabbit} \\
Offering more food influences neither $X$ nor $Y$.  Adding or removing rabbits changes $X$, but not $Y$, while the appetizer changes both $X$ and $Y$
and preserves the equality $X = n \cdot Y$. 
We thus have actions influencing both (while preserving their relation) and those influencing only $X$,  but no action influencing only $Y$. We thus conclude $Y \to X$.

\vspace{0.1cm}
\noindent
{\bf Scenario 2: shortage of food}\\ 
Now, the appetizer has no effect. Changing the number of rabbits changes the food per rabbit, but not the total consumption. 
Changing the amount of food changes consumption per rabbit and total consumption. Hence we have actions that 
influence $Y$, but not $X$, and actions that influence both, but preserve the relation $Y=X/n$     
Further, we have no action influencing $X$ without affecting $Y$. We thus conclude $X \to Y$.
\end{Example}

The above dependence of the causal direction on the regime suggests that there exists a large grey zone without clearly defined causal direction.
We want to discuss this for the following business relevant example, where we will not offer a clear answer. 
\begin{Example}[revenue and sold units]\label{ex:ru}   
Let $R$ and $Q$ be random variables denoting the revenue and the number of sold units for a company, respectively.
Its instantiations are $r_j,q_j$ denoting revenue and number of units for product $j$. If $p_j$ denotes the price of product $j$ we thus have
\begin{equation} 
r_j = p_j \cdot q_j.
\end{equation} 
If we consider $n_j$ as instantiations of a random variable $N$, we thus write
\begin{equation}\label{eq:ru}  
R = P \cdot Q.
\end{equation} 
One may want to read \eqref{eq:ru} as structural equation, which suggests the causal structure $Q\to R$. 
However, this requires $P$ to be independent of $Q$  as a minimal requirement (unless we are talking about a confounded relation).
This independence is often violated  when more items are sold for {\it cheap} products.
On the other hand, we cannot expect $P$ to be independent of  $R$ either, hence neither \eqref{eq:ru} nor $Q= R/P$ should be considered a structural equation.

To argue that $Q$ causes $R$ one may state that a marketing campaign can increase the revenue by increasing the number of sold units.
However, why is a marketing campaign an intervention on $U$ rather than on $R$ if it increases both by the same factor? 

While our intuition may consider $U\to R$ as the `true' causal direction, the following scenario challenges this. 
Assume there are two farmers, farmer $P$ producing potatoes, and farmer $E$ producing eggs. They have implemented a countertrade with exchanging $K_P$ and $K_E$ many potatoes and eggs, respectively, according to the negotiated exchange factor $F$. We then have
\begin{equation}\label{eq:farmers} 
K_E = K_P \cdot F.
\end{equation} 
For farmer $P$, $K_P$ is the number of units sold while $K_E$ is the revenue, while farmer $E$ considers $K_E$ the number of units sold and $K_P$ the revenue. If number of units is always the cause of the revenue, then the causal direction depends on the perspective. Preference for one causal direction versus the other could come from insights about which quantity reacts more to changes of $F$: 
assume, for instance, the number of potatoes exchanged is more robust to changes of $F$ (i.e., in economic terms, the demand of potatoes has small price elasticity), we could consider  changes of $F$ as interventions on $K_E$  and thus conclude $K_P \to K_E$. 
\end{Example} 
Example \ref{ex:ru} shows that causal directions can also be in grey zones because {\it actions} can be in grey zone of 
being an intervention on one versus the other quantity. 
Acting on the factor $F$ will in general 
change both variables $R$ and $U$. However, in the regime where one of it is relatively robust to changes of $F$, we can consider this one as the cause and 
consider changing $F$ as an intervention on the effect because it affects the mechanism relating cause and effect.  
It is likely that many causal relations in real life show equally much room for interpretations. 

Let us now revisit the example from the domain of product recommendation algorithms described in \cite{Schoelkopf2021}:
\begin{Example}[Laptop and its rucksack]\label{ex:laptop}
Let $X,Y$ be binaries that describe the decisions of a person to buy or not to buy a laptop and a laptop rucksack, respectively.
Let $P(X,Y)$ be the prior joint distribution without any marketing campaign. Let actions in $\cA_l$ be  marketing campaigns that try to sell more
laptops (without explicitly mentioning laptop rucksacks). Then it is likely that these  actions influence $P(X)$, but not $P(Y|X)$.  
Let actions $\cA_r$ define marketing campaigns that target at selling laptop rucksacks. Let us assume that this changed
 $P(Y|X)$, but not $P(X)$. 
 \end{Example} 
Here we have neglected that seeing laptop rucksacks may remind some customers that they were planning to buy a laptop already since a while, which could induce 
additional demand for laptops. Further, a marketing campaign changing $P(X)$ and $P(Y|X)$ could certainly exist, e.g., one that 
explicitly advertises laptop and rucksack as an economically priced pair.    
We are thus aware of the fact that any causal statement in this vague domain of
customer psychology is a good approximation at best. 
When previously mentioning the scenario of Example \ref{ex:laptop} in the introduction we have  emphasized that the {\it time order} of customer's purchases
does not determine the {\it causal order} of the underlying decisions for the purchases.
This already suggests that analyzing the causal order of the underlying materialized processes does not reveal the causal structure
of their psychologic origin. The following example elaborates on this discrepancy of phenomenological causal direction and causal direction of unerlying
micro-processes.   
\begin{Example}[vending machine]\label{ex:vending} 
In contrast to usual purchasing process, a vending machine outputs the article clearly {\em after} and {\em because} the money has been inserted. Accordingly,
inserting the money is the cause of obtaining the product. We will call this causal relation `microscopic'.
For some cigarette vending machine, let $X$ be the number of packages sold at a day and $Y$ be the total amount of money inserted at the same day. 
Our microscopic causal relations suggests to consider $Y$ the cause of $X$, but previous remarks on the relation between revenue and number of sold units
suggest the opposite. Let us therefore ask for `natural actions' on the system. 
Assume we stop some of the smokers on their way to the machine and convince them not to buy cigarettes. This clearly impacts both $X$ and $Y$. 
Another action would be to slightly change the price of the packages by manipulating the vending machine. If the change is small enough, it will only
affect $Y$ but not $X$. We thus have a natural action influencing both and one influencing only $Y$, which suggest that $X$ influences $Y$, in agreement with what we said about revenue and sold units, but in contrast to the {\em microscopic} causal structure. 
\end{Example} 

\subsection{The multivariate case} 
We first generalize Definition \ref{def:phcestat} to multiple variables. Although these generalizations are straightforward, 
we will see that our multivariate extensions of the urn example reveal the abstractness of phenomenological causality even more. 
\begin{Definition}[multivariate causality, statistical]\label{def:phenDAGstat} 
Let $\cA$ be elementary actions on a system described by  the variables $X_1,\dots,X_n$. Then we say that $G$ is a valid causal graph
if $\cA$ consists of classes $\cA_1,\dots,\cA_n$ such that actions in $\cA_j$ change no other conditional than that $P(X_j|PA_j)$.
\end{Definition} 
Likewise, we generalize Definition \ref{def:pheceunit}:
\begin{Definition}[multivariate causality, unit level]\label{def:phenDAGunit}  
Adopting the setting from Definition \ref{def:phcestat} we say that $G$ is a valid causal graph if $\cA$ decomposes into classes $\cA_j$ such that 
for every statistical instantiation $(x_1,\dots,x_n)$ there are maps $m_1,\dots,m_n$ with 
\begin{equation}\label{eq:fcmunit}   
x_i = m_i(pa_i),
\end{equation} 
such that actions in $\cA_j$ preserve all equations \eqref{eq:fcmunit} valid for $i\neq j$. 
\end{Definition} 

We now generalize Example \ref{ex:urn} to $n$ different balls, where the causal structure suggested by our definition gets 
even less obvious:
\begin{Example}\label{ex:nballs} 
Given $n$ balls with labels $j=1,\dots,n$. Given the actions $A^+_j$  and $A^-_j$, 
which replace one ball of type $j-1$ with $j$ for $j=2,\dots,n$ or vice versa, respectively. Further, $A_1^\pm$ are 
defined as adding or removing
balls of type $1$. If $k^0_j$ denotes the initial number of balls of type $j$, and $N_j$ denotes the number of actions
$A^+_{j}$ minus the number of $A^-_j$, the number $K_j$ of balls is given by
\begin{eqnarray} \label{eq:urns}
K_j &=& k^0_j + N_{j} - N_{j-1} \quad \hbox{ for } j \geq 2\\
K_1 &=& k_n^0 + N_{1}. \label{eq:urns0}
\end{eqnarray}
Let us first recalibrate $K_j$ to $\tilde{K}_j:= K_j - k^0_j$.  
We then introduce vectors $\bk^0:=(k^0_1,\dots,k^0_n)$ and vector valued variables $\tilde{\bK}:=(\tilde{K}_1,\dots,\tilde{K}_n)^T$, $\bN:=(N_1,\dots,N_n)^T$.
Using the Töplitz matrix
$S$ with diagonal $1$ and second diagonal $-1$ (and zero elsewhere), we can rewrite \eqref{eq:urns} and \eqref{eq:urns0} as   
\begin{equation}\label{eq:urnica} 
\tilde{\bK} = S \bN.
\end{equation}  
This, in turn, can be rewritten as
\begin{equation}\label{eq:urnAdef}
\tilde{\bK} = A \tilde{\bK} +\bN,
\end{equation}   
with the lower triangular matrix
\begin{equation}\label{eq:urnAder} 
A:=I - S^{-1} =  \left(\begin{array}{ccccc} 0 &  &\cdots    & & 0 \\ -1 & 0 &  &  &  \\ \vdots & -1 & \ddots &  & \vdots \\   & &  & &  \\ 
-1&  & \cdots & -1 & 0\end{array}\right).   
\end{equation} 
Equivalently, we can then rephrase \eqref{eq:urnica}  by the structural equations
\begin{eqnarray}
\tilde{K}_j &=& \sum_{i > j} - \tilde{K}_i + N_j. \label{eq:Kseq}
\end{eqnarray}
The causal structure for the $K_j$, which is the same as for $\tilde{K}_j$,  is shown in Figure \ref{fig:doublechain}. 
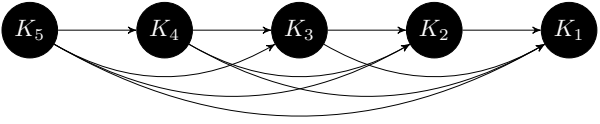
\begin{figure} 
\begin{center}
\resizebox{8cm}{!}{%
\begin{tikzpicture}
 \node[obs] at (-5,0) (K5) {$K_5$} ;
 \node[obs] at (-3,0) (K4) {$K_4$} edge[<-] (K5) ;
 \node[obs] at (-1,0) (K3) {$K_3$} edge[<-] (K4) edge[<-, bend left] (K5);
 \node[obs] at (1,0) (K2) {$K_2$} edge[<-] (K3) edge[<-, bend left] (K4) edge[<-, bend left] (K5) ;
 \node[obs] at (3,0) (K1) {$K_1$} edge[<-] (K2) edge[<-, bend left] (K3) edge[<-, bend left] (K4) edge[<-, bend left] (K5)  ; 
\end{tikzpicture} 
} 
 \end{center} 
\caption{\label{fig:doublechain} Causal relation between the variables $K_j$, which count the number of balls in the urn with label $j$ (according to our definition of phenomenological causal structure).} 
\end{figure} 
Note that there is exactly one structure matrix $A$ that 
admits writing each $K_j$ as a linear expression of some $K_i$ and $N_j$ such that $A$ is lower triangular for some ordering of nodes.
This is because $S$ uniquely determines $A$. 
Assuming linear structural equations, we thus obtain Figure \ref{fig:doublechain}  as the unique DAG corresponding to the 
defined set of elementary actions.
\end{Example}
Note that the algebraic transformations between \eqref{eq:urnica} and \eqref{eq:urnAder}  resemble the algebra in 
 Independence Component Analysis (ICA)-based  multivariate causal discovery  \cite{Moneta} (following the idea of LiNGAM \cite{Kano2003} mentioned 
 for the bivariate case above). This analogy is not a coincidence: ICA decomposes the vector $\bK$ into independent noise variables
 $\bN$. Accordingly, since \eqref{eq:urnAdef} is a linear acyclic causal model with {\it independent non-Gaussian} noise variables $N_j$, multivariate LiNGAM would also identify the same causal structure and FCMs that we derived as phenomenological causal model. 
 In other words, if we ensure that the choice of the actions is controlled by random generators, independently across different $\cA_j$, we obtain
 a joint distribution $P(K_1,\dots,K_n)$ for which the causal discovery algorithm LiNGAM infers the DAG  in Figure \ref{fig:doublechain}. 

It is instructive to discuss Example \ref{ex:nballs} from the perspective of complexity of some actions that are {\it not} elementary.
Increasing $K_j$ without affecting the others requires $j$ operations, e.g.,one can first increase $K_1$ and propagate this increase to $K_j$. From the causal perspective, these actions are necessary to compensate the impact of $K_j$ on its child.    

A further remark on causal faithfulness \cite{Spirtes1993}. The fact that an intervention only propagates to the child, but not to the grandchild
shows that the structural equations are non-generic; direct and indirect influence of $K_j$ on $K_{j-2}$ compensate. 
Accordingly, if we control each action by independent coin flips as in the remarks after Example \ref{ex:urn},   
the induced joint distribution 
will not be faithful to the causal DAG. The idea of `nature choosing each mechanism $p(x_j|pa_j)$ in \eqref{eq:fac} independently' seems to have its limitation here.
The reason is that the actions $A_j^\pm$ are the building blocks of the system, rather than the Markov kernels $p(x_j|pa_j)$, which are constructed
from the former. 
There is also another `paradox' of our causal interpretation that becomes apparent for $n>2$, while it seems less paradoxical in Example \ref{ex:urn}:
imagine what happened if we were to redefine $A_0^\pm$ as adding or removing of balls of type $n$ instead of type $1$. We would then reverse all the arrows
in Figure \ref{fig:doublechain}. In other words, the direction of the arrows in a long chain of variables depends on what happens {\it at the end points}.
This idea is in stark contradiction to the spirit of modularity \cite{HauWoo99} assuming each $p(x_j|pa_j)$ is an independent mechanism of nature. 
The reader may see this as an indicator against interpreting the equations \eqref{eq:Kseq} as FCMs, but we think that causal directions on the phenomenological level
may well depend on this kind of {\it context}.

In Example \ref{ex:nballs} the locality of the impact of each of the actions $A^\pm_j$ itself (affecting only $2$ adjacent variables) entailed long-range causal influence
between the variables. 
Now we will describe the opposite where actions affecting a large number of variables is induced by only {\it local}  causal connections
(in other words: in the first example $S$ has only entries in the first off-diagonal, in the case following now this is true for $A$).  
\begin{Example}[$n$ different balls in bundles]\label{ex:bundles} 
We now modify Example \ref{ex:nballs} such that the $n$ balls come in the following bundles: there are $n$ different types of packages and type $P_j$ contains 
the balls $1,\dots,j$ (one per package). Then there are $2n$ different actions $A^+_1,A^-_1,\dots,A^+_k,A^-_k$
of the following form: $A_j^+$ puts one package $P_j$ from the stack into the urn, while $A_j^-$ wraps balls with label $1,\dots,j$ to one package and puts them back to the stack. 
We then introduce $n$ random variables, $K_1,\dots,K_n$, where $K_j$ 
is the number of balls with label $j$ in the urn. Obviously transformation $A^+_j$ simultaneously increases 
all the variables $K_1,\dots,K_j$ by $1$, while $A^-_j$ decreases all of them by $1$, as depicted in Figure \ref{fig:ballchain} for $n=4$. 
\begin{figure}
\centerline{
\includegraphics[width=0.48\textwidth]{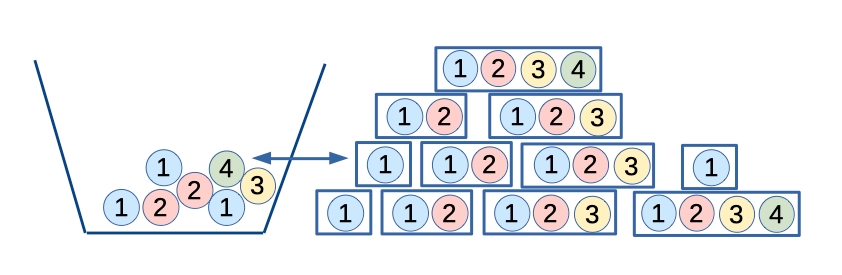} 
} 
\caption{\label{fig:ballchain} Urn containing packages of $n$ different types, where type $P_j$ contains
balls will label $1,\dots,j$. The variable $K_j$ counts the number of balls with symbol $j$ in the urn. 
The elementary operations of the system are adding one package from the stack to the urn or put it back. 
Changing $K_j$ thus entails the same change for $K_{j-1},\dots,K_1$.} 
\end{figure} 

Using the same derivation and notation as in Example \ref{ex:nballs}, we define $N_j$ as the difference of actions $A_j^\pm$ and obtain
\begin{equation} 
\tilde{K}_j = \sum_{i \geq j} N_i, 
\end{equation}  
which yields $\tilde{\bK}= S \bN$ with 
\[
S:= \left(\begin{array}{ccccc} 1 & 0 &\cdots    & & 0 \\  & 1 & 0 &  &  \\ \vdots &  & \ddots &  & \vdots \\   & &  & &  \\ 
1&  & \cdots &  & 1\end{array}\right)
\] 
For the structure matrix, we thus obtain the lower triangular matrix 
\[
A = I - S^{-1} =  \left(\begin{array}{ccccc} 0 &  &\cdots    & & 0   \\ 1 & 0 &  &  &  \\   0 & 1 &  &     & 0  \\ \vdots &   & \ddots &  & \vdots \\   & &  & &  \\ 
0&  \cdots &0 & 1 & 0\end{array}\right),
\] 
which amounts to
the structural equations
\begin{eqnarray}
K_n &=& N_n,\\ \label{eq:seballchain1} 
K_j &=& K_{j+1} + N_j \quad \forall j \leq n-1. \label{eq:seballchain2}.
\end{eqnarray} 
These equations correspond to the causal DAG in Figure \ref{fig:chain}. 
\begin{figure} 
\begin{center}
\begin{tikzpicture}
 \node[obs] at (-3,0) (K4) {$K_4$} ;
 \node[obs] at (-1,0) (K3) {$K_3$} edge[<-] (K4);
 \node[obs] at (1,0) (K2) {$K_2$} edge[<-] (K3);
 \node[obs] at (3,0) (K1) {$K_1$} edge[<-] (K2); 
\end{tikzpicture} 
 \end{center} 
\caption{\label{fig:chain} Causal relation between the variables $K_j$, which count the number of balls in the urn with label $j$ (according to our definition of phenomenological causal structure).} 
\end{figure}
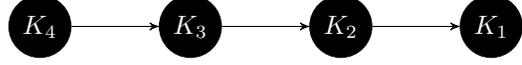 
An intervention that changes $K_j$ necessarily changes all $K_s$ with $s< t$ 
by the same amount, as a downstream impact, according to \eqref{eq:seballchain2}. 
While the transformations $A^\pm_j$ change all $K_s$ with $s<t$ per definition, 
it is a priori not obvious to see which of these changes should be considered direct and which one indirect.
However, the causal interpretation  \eqref{eq:seballchain2} clearly entails such a distinction. 
\end{Example} 

\paragraph{What's the purpose of the causal interpretation?} 
The balls in the urn show an extreme case where the causal interpretation is far away from any `mechanistic view' of causality where the functions 
$m_j$ from Definition \ref{def:phenDAGunit} refer to tangible mechanisms (recall, for instance that $\cA_j$ in Example \ref{ex:nballs} were symmetric with respect to swapping $j$ and $j-1$, yet we have identified them as interventions on $K_j$, not on $K_{j-1}$). To argue that our causal interpretation is not just a weird artifact of our concept, we need to show its benefit. The following section will argue that this extension of causality allows us to consistently talk about the overall causal DAG
when systems with `phenomenological causality' are embedded in systems with more tangible causal structure. 

\paragraph{Related ideas in the literature}
While the idea that causal conditionals $p(x_j|pa_j)$ and structural equations 
define `modules' 
in a causal DAG which can be manipulated independently 
has a long tradition, we have turned it somehow around by defining these manipulations as the primary concept and the causal DAG (in case the set of 
elementary actions correspond to a DAG, see below) as a derived concept. The closest work to this idea seems to be \cite{Blom2021}, where DAGs also appear as derived concepts rather than being primary. The idea is to start with a set of equations, where each can contain containing endogenous variables as well as exogenous ones. Subject to certain conditions (phrased in terms of matchings in bipartite graphs), one can uniquely solve the equations to express the endogenous variables in terms of the exogenous ones using Simon's ordering algorithm \cite{Simon1953}.  Remarkably, general interventions in \cite{Blom2021} are thought to act on {\it equations}  rather than {\it variables}. Note that in the usual view of causality, an action that changes the structural equation $X_j = f_j(PA_j,N_j)$ to some different
equation $X_j = \tilde{f}_j(PA_j,N_j)$ is considered an intervention {\it on} $X_j$ only because the equation is read as a ``structural equation'' 
(or ``assignment'') for
$X_j$, rather than  for any of the parents or the noise $N_j$. In \cite{Blom2021,Simon1953}, an equation is not {\it a priori} considered an assignment for 
a certain variable, but only later after analyzing the direction in which the system of equations is solved. 
This way, causal direction also emerges from the context of the entire set of equations, while ours emerges from the context of other actions. 
However, even this difference is less substantial than it appears at first glance. After all, sets of equations can be turned into equivalent sets of 
equations, but in the different set of equations changes of one equation may translate into changes of many equations.
Therefore we assume that in  \cite{Blom2021} the preference for any of these equivalent set of equations comes from an implicit notion of {\it which changes 
of the system are more elementary than others}. 
The question where this notion of complexity of actions come from goes beyond the scope of this paper. We hope that 
Examples like Ex.\ref{ex:ball_track} showed that in real life scenarios there are reasons to consider some actions as obviously 
more elementary than others. Further, we refer to the appendix where we argue that complexity of actions can be
subject of scientific research and mention some approaches from modern physics.

\section{Phenomenological causality couples to tangible causality \label{sec:coupling}}
One can argue that a crucial property of causality is to describe the way a system with some variables couples to other variables in the world. 
In \cite{Tsamardinos,janzing2018merging,Gresele2022}, causality is used to predict statistical relations of variables  that have not been observed together. 
This section shows in which sense a causal DAG defined via phenomenological causality can be consistently embedded into the context of further variables. 
The mathematical context of the below observations is fairly obvious and mostly known. 
Yet, we consider them crucial as justification of phenomenological causality.

\subsection{Markov property of phenomenological causality \label{subsec:MarkovExt}} 
Let us first consider the mechanisms described by functions $m_j$ in Definition \ref{def:phenDAGunit}. Since they represent  structural equations
$f_j(.,n_j)$ with fixed noise value $n_j$, we will denote them with superscript and write $m_j^{n_j}$. Whenever the noise values $n_j$ are statistically independent
across different statistical units, they induce a joint distribution that is Markovian with respect to $G$, see \cite{Pearl:00}, Theorem 1.4.1. 
We conclude that we obtain a Markovian joint distribution of $P(X_1,\dots,X_n)$ whenever we control actions in $\cA_j$ by independent random variables.
The same holds true when we control the actions in Definition \ref{def:phenDAGstat} by independent random variables and introduce 
formal random variables $\Theta_j$ controlling the causal conditionals $p^{\theta_j} (x_j|pa_j)$. Then the joint distribution 
\begin{eqnarray*}
&& P(X_1,\dots,X_n) \\
&=&  \int \prod_{j=1}   p^{\theta_j} (x_j|pa_j) p(\theta_1)\cdots p(\theta_n) d\theta_1 \cdots d\theta_n \\
&=& \prod_{j=1}^n  \int p^{\theta_j}  (x_j|pa_j) p(\theta_j) d\theta_j, 
\end{eqnarray*} 
still factorizes with respect to $G$. Note that the assumption of independent  $\Theta_j$ is in agreement with how \cite{Zhang2017CausalDF} interpret
the postulate of independent mechanisms (see e.g. \cite{causality_book}, Section 2.1), namely as {\it statistically independent changes} of the causal conditionals $p(x_j|pa_j)$ across environments. 
While \cite{Zhang2017CausalDF} use this property for identification of the causal DAG, we use it to show that then the distribution averaging over different environments is still Markovian.
In other words, $G$ is true both with respect to each environment but also with respect to the aggregated distribution. 
Moreover, for linear structural equations, as for instance, \eqref{eq:urnAdef},  also the causal discovery method LiNGAM would infer a causal structure that aligns with
phenomenological causality, as mentioned earlier. 

A more interesting scenario, however, is obtained when elementary actions are controlled by further random variables $Y_1,\dots,Y_m$ which are connected by a non-trivial causal structure.
We argue that then we obtain a joint distribution on $X_1,\dots,X_n,Y_1,\dots,Y_m$ whose DAG is consistent with phenomenological causality. 
Assume, for instance, that some actions are not only controlled by independent noise variables $N_j$ or $\Theta_j$, respectively, but by 
one of the variables $Y_i$ which are related by a DAG themselves. We then model the influence of $Y_i$ on actions in $\cA_j$ by 
introducing a second superscript to the mechanisms $m_j$ and $p(x_j|pa_j)$, respectively, and obtain $m^{y_i,n_j}_j$ or $p^{y_i,\theta_j}(x_j|pa_j)$.
Obviously, this way $Y_i$ can be read as an additional parent of $X_j$.
Further, let some $Y_l$ be influenced by some $X_j$ by modifying the structural equations for some $Y_l$ such that they receive $X_j$ as additional input.  

We now define a directed graph with nodes  $X_1,\dots,X_n,Y_1,\dots,Y_m$  
by drawing an edge from $Y_i$ to $X_j$ whenever $Y_i$ controls actions in the set $\cA_j$ and draw an edge from $X_j$ to $Y_l$ whenever the latter is influenced by the former. Whenever this graph is a DAG $\tilde{G}$, $P(X_1,\dots,X_n,Y_1,\dots,Y_m)$   will clearly be Markovian relative to $\tilde{G}$. 
This is because the generating process, by construction, follows structural equations according to $\tilde{G}$ and the joint distribution admits a corresponding Markov factorization. 

\begin{figure} 
\includegraphics[width=0.5\textwidth]{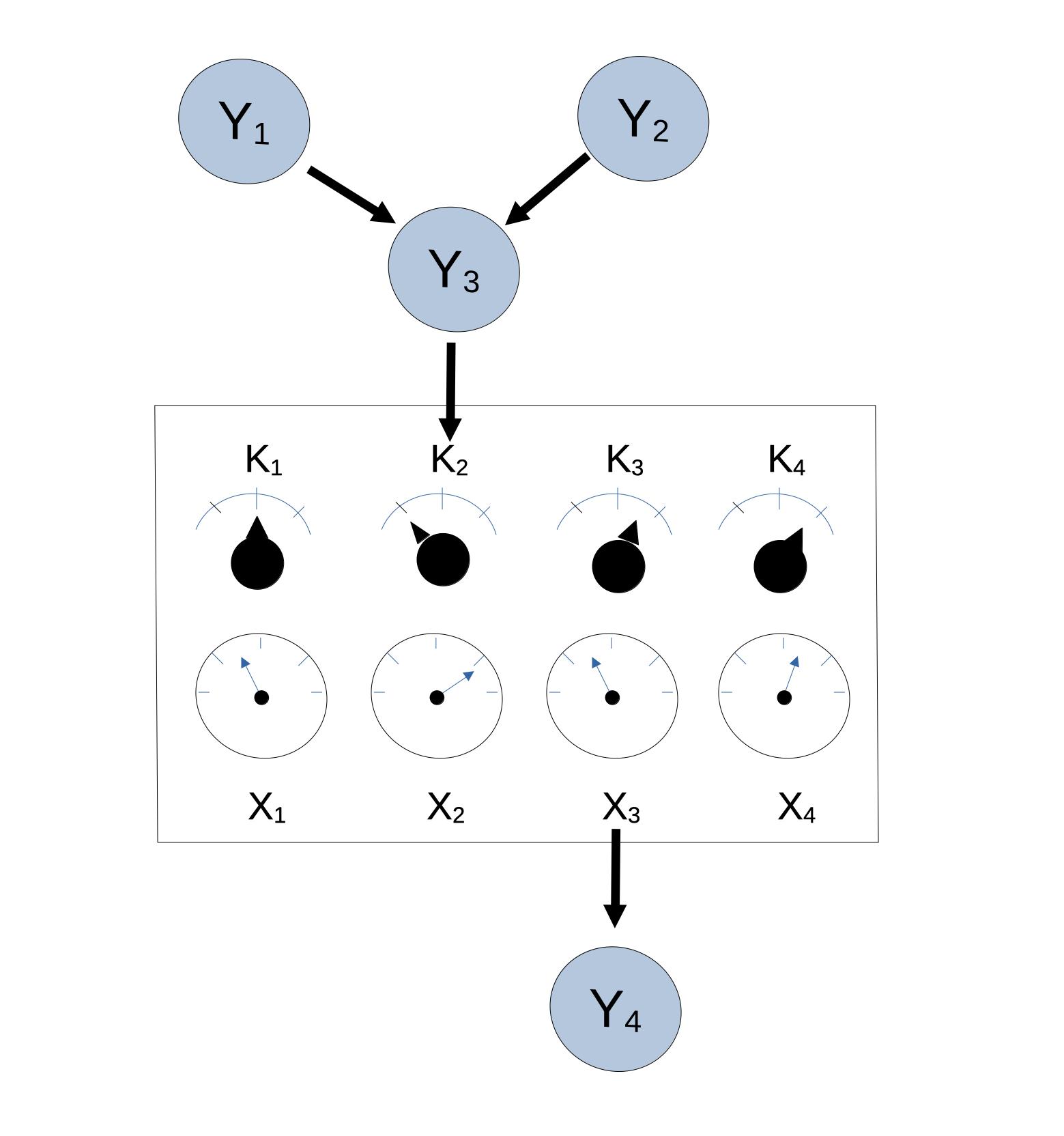} 
\caption{\label{fig:boxtangible} Visualization of a scenario where phenomenological causality couples to variables with tangible interventions.
Then our construction of abstract causal relations  between the variables $X_j$ are justified by consistency in the sense of a Markov condition
for the causal DAG of the joint system.
} 
\end{figure}

In a scenario where causal relations among the $Y_i$  and among $Y_i$ and $X_j$ are justified by tangible interventions, the abstract notion of causality between
different $X_j$ thus gets justified because it is consistent with the causal Markov condition also after embedding our abstract system into the tangible world. 
Getting back to our metaphor with a box with $n$ knobs and $n$ displays, our phenomenological definition of the causal relations inside the box is consistent
with the DAG that describes causal relations between the box and the more tangible world, see Figure \ref{fig:boxtangible} for a visualization.  
Since the causal structures for our Examples \ref{ex:urn}, \ref{ex:nballs}, and \ref{ex:bundles} seemed particularly artificial, 
it arguably gets less artificial once such a system gets embedded into an environment with further variables.

\subsection{Boundary consistency of the notion 'elementary' \label{subsec:boundary}}
We emphasized  that determining whether a  variable $X_j$ is `directly' affected by an action (and thus deciding whether the action is an intervention on
$X_j$) is a causal question that may be equally hard to decide as the causal relations between the variables $X_1,\dots,X_n$.
Formally, the question can be phrased within a meta-DAG containing an additional variable $A$ describing the actor, in which the relation between $A$ and 
$X_j$ appears as a usual link if and only if $A$ is an intervention on $X_j$. 
Due to this arbitrariness of the boundary between system and actor  we expect a framework 
for causality to be consistent with respect to shifting this boundary (by extending or reducing the set of variables).\footnote{In the early days of quantum physics, Heisenberg described a similar consistency of the theory with respect to shifting the boundary between {\it measurement apparatus} and {\it quantum system  to be measured} (the `Heisenberg cut'). Ref. \cite{PhysUniversal} is even more similar in spirit to our boundary consistency because it describes the arbitrariness of the boundary between {\it controlling device} and {\it system to be controlled} for interventions on microscopic physical systems and constructs a framework for physical controllers in which this boundary can be shifted in a consistent way.} 
Here we want to backup our notion of 'elementary' by the argument that 
it is consistent with respect to a certain class of marginalizations to subsets of variables. 
To explain this idea, we first introduce a rather strong notion of {\it causal sufficiency}:\footnote{Note that this notion has been introduced as 'causal sufficiency'  in \cite{causality_book}, but the sentence in the bracket has been forgotten, as 
 noted in the errata of the book.}  

\begin{Definition}[graphical causal sufficiency] 
Let $\bX:=(X_1,\dots,X_n)$ be nodes of a DAG $G$. A subset $\bX_S$ is called graphically causally sufficient 
 if there is no hidden common cause $C \notin \bX_S$ that is causing at
least two variables in $\bX_S$ (and the causing paths go only through
nodes that are not in $\bX_S$).
\end{Definition}

In general, the model class of causal DAGs is not closed under marginalization, but requires the model class of Maximal Ancestral Graphs (MAGs) \cite{Richardson2002}. Here we restrict the attention to the simple case of graphical causal sufficiency, where the causal model remains in the class of DAGs after marginalization: 
\begin{Definition}[marginal DAG]
Let $\bX$ be the nodes of a DAG $G$ and $\bX_S$ a graphically causally sufficient set. Then the marginalization $G_S$ of $G$ to the nodes $\bX_S$ 
is the DAG with nodes $\bX_S$ and an edge $X_i\to X_j$ whenever there exists a directed path from $X_i$ to $X_j$  in $G$  containing no node from $\bX_S$
(except $X_i,X_j$).
\end{Definition} 
To justify the definition, we need to show that the distribution of $\bX_S$ is Markov relative to $G_S$ and that $G_S$ correctly describes interventional probabilities. 
It is easy to check the Markov condition: Let $\bX_A,\bX_B,\bX_C$ be subsets of $\bX_S$ such that
$\bX_A$ is $d$-separated from  $\bX_B$ by $\bX_C$ in $G$, hence every path in $G$ connecting a node in $\bX_A$ with one in $\bX_B$
contains either (i)  a chain or a fork with middle node in $\bX_C$ or (ii) an inverted fork whose middle node is not $\bX_C$ and also not its descendants.
It is easy to see that conditions (i) and (ii) are preserved when directed paths are collapsed to single arrows, and thus the same conditions hold in
$G_S$. To see that interventions on arbitrary nodes in $\bX_S$ can equivalently be computed from $G_S$, we recall that 
interventional probabilities can be computed from backdoor adjustments \cite{Pearl:00}, Equation (3.19). 
We can easily verify that if
$Z\subset \bX_S$ satisfies the backdoor criterion in $G_S$ relative to an ordered pair $(X_i,X_j)$ of variables in $\bX_S$, it also satisfies it in $G$ because the property of blocking backdoor paths is inherited from $G$.

The following result shows that our notion of `elementary' is preserved under  marginalization to causally sufficient subsets: 
\begin{Theorem}[boundary consistency]  
Let  $G$ be a DAG with nodes $\bX:=\{X_1,\dots,X_n\}$ and 
$P,\tilde{P}$ be joint distributions of $\bX$ that are Markov relative to $G$ and differ only by one term in the factorization \eqref{eq:fac}. 
For some subset $S$ of nodes satisfying graphical causal sufficiency, 
let $G_S$ with $\bX_S\subset \bX$ be a marginalization of $G$, and $P_S,\tilde{P}_S$ be marginalizations of $P,\tilde{P}$, respectively.
Then $P_S$ and $\tilde{P}_S$ also differ by one conditional at most. 
\end{Theorem} 
\begin{Proof} Let $P$ and $\tilde{P}$ differ by  the conditional corresponding to $X_j$. 
Introduce a binary variable $I$ pointing on $X_j$ which controls switching between $P(X_j|PA_j)$ and $\tilde{P}(X_j|PA_j)$. Formally, we thus define
a distribution $\hat{P}$ on $(\bX,I)$ 
such that $\hat{P}(x_j|pa_j, I=0) = P(x_j|pa_j)$ and  $\hat{P}(x_j|pa_j, I=1) = \tilde{P}(x_j|pa_j)$.
Let $G^I$ be the augmented DAG containing the nodes of $G$ and $I$ with an arrow from $I$ to $X_j$.  
For the case where $X_j\in \bX_S$, it is sufficient to show that the marginalization of $G^I$ to $S\cup \{I\}$ does not connect  $I$ with any node $X_i$ other than $X_j$, which follows already from the fact that any directed path from $I$ to $X_i$ passes $X_j$. 
Now assume that $X_j$ is not in $\bX_S$. By causal sufficiency of $\bX_S$, there is a unique node $X_{\tilde{j}}$ among the 
descendant of $X_j\in \bX_S$ that blocks all paths to other nodes in $\bX_S$  (otherwise $\bX_S$ would not be causally sufficient).
Hence, the DAG $G_S^I$ contains only an edge to   $X_{\tilde{j}}$ but no other node in $\bX_S$.
 \end{Proof}

\begin{figure}    
\centerline{
\resizebox{3cm}{!}{%
\begin{tikzpicture}
    \node[obs] at (0,-3) (X3) {$X_3$} ; 
    \node[obs] at (0,-1.5) (X2) {$X_2$}  edge[->] (X3) ;
    \node[] at (2,-1.5) (I) {} edge[->] (X2);
    \node[obs] at (0,0) (X1) {$X_1$} edge[->] (X2) ; 
\end{tikzpicture}  
}   
\hspace{1cm} 
\resizebox{3cm}{!}{%
\begin{tikzpicture}
    \node[obs] at (0,-3) (X3) {$X_3$} ; 
    \node[] at (2,-1.5) (I) {} edge[->] (X3);
    \node[obs] at (0,0) (X1) {$X_1$} edge[->] (X3) ; 
\end{tikzpicture}  
}
}
\caption{\label{fig:marginalization} Left: action on node $X_2$, which results in an action on $X_3$ after dropping node $X_2$ (right).} 
\end{figure}
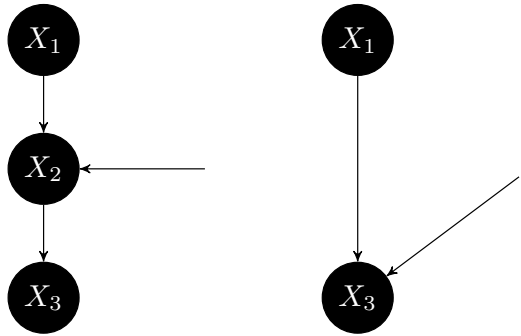
It seems that every framework that is supposed to be general enough to describe significant aspects of the world should not only be able to
describe the system under consideration, but also its interaction with agents.  
Understanding why and in what sense certain actions are more elementary than others is still a question to be answered outside the framework.
However, demanding consistency of different boundaries 
between system and intervening agents seems a more modest and feasible version of `understanding' of how to define elementary. 

\section{Conclusions}
We have described several scenarios --some of them are admittedly artificial, but some of them are closer to real-life problems--   where causal relation between observed quantities are not defined {\it a priori}, but get only well-defined after specifying the `elementary actions'  that are considered interventions on the respective variables. 
We have argued that this specification admits the definition of 
an abstract notion of causality in domains where the mechanistic view of tangible causal interactions fails. 
We believe that this approach renders the context-dependence of causality more transparent since there may be different {\it elementary actions} in different contexts.
It is possible that at least some part of the fuzziness  of some relevant causal questions (e.g. `does income influence life expectancy?') 
comes from the missing specification of actions. From this point of view one could argue to accept only causal questions
that directly refer to the treatment effect for which {\it the treatment itself} is obviously a feasible action (e.g. taking a drug or not) and rejecting
questions about the causal effect of variables like `income'. However, our approach is different in the sense that --after having defined the elementary actions-- it
does talk about causal relations between variables `inside the box of abstract variables', that is, variables for which interventions are not defined a priori.
This is because we believe that analyzing causal relations `inside the box' is crucial for understanding complex system.  



 
\appendix






\paragraph{Acknowledgements} 
Many thanks to Joris Mooij for inspiring discussions on the relation to \cite{Blom2021}.


\appendix

\section{\label{sec:physics} Complexity of actions in modern micro-physics}

Notions of complexity of transformations have traditionally been subject of computer science in the sense of {\it computational} complexity.
In a nutshell, computational complexity explores how the number of {\it elementary logical operations} scales with the problem size. 
While complexity theory does not come with an advice which logical transformations are supposed to be elementary, the asymptotic scaling
behaviour is independent of this convention provided that they can be defined as operations of a universal Turing machine \cite{Papadimidriou2003}. 
Computer science has therefore considered different models of computation as basis for complexity theory. 
While asymptotic behaviour can be a good heuristic to estimate running time for real problems, the question where a notion of complexity 
in our finite world should come from remains actually open. 

However, there are ideas from modern physics that provide new insights in this regard. 
The preceding three decades of quantum information research \cite{NC} has intertwined 
computer science and physics in a way that the disciplines have never seen before. First, \cite{Deutsch} emphasized 
that {\it the laws of physics} determines which logical operations are simple and complex and argued that the laws of quantum physics
entail a new notion of {\it Quantum Complexity} -- which may differ from complexity of classical computer science-- but seems more fundamental since 
it is a notion of complexity that is defined via microscopic physical processes. There, logical operations are considered elementary because 
one can describe existing physical interactions that implement them. 
Second, the elementary operations in Quantum Computing \cite{NC} 
can not only be interpreted as {\it logical} operations, but also as operations whose goal is more general than only implementing a computation.
Researchers considered, for instance, the complexity of {\it cooling algorithms}, that is, complex transformations on
molecular systems that transfer heat from one part to the other \cite{Fernandez}, and similarly, the complexity of
{\it heat engines} \cite{HeatEngines}. Further, several articles considered the complexity of measurement processes
 \cite{PSPACE_QIC,Yosi2017} and compared their complexity with  hard {\it computational} tasks (see also \cite{Habil} 
for a slightly outdated overview). The essential message for the present paper is that complexity of actions is indeed, despite the fuzziness of the question, 
subject of scientific research.

\end{document}